\documentclass[11pt,a4paper]{article}

\title{PRANK: a singular value based noise filtering of multiple response datasets for experimental dynamics}
\author{\textbf{F. Trainotti $^1$, S. W. B. Klaassen $^2$, T. Bregar $^3$, D. J. Rixen $^1$} \\
	$^1$ Technical University of Munich, Department of Mechanical Engineering, \\ Chair of Applied Mechanics, Boltzmannstr. 15, 85748 Garching, Germany\\
	e-mail: \textbf{francesco.trainotti@tum.de}, \textbf{rixen@tum.de} \\
	$^2$ Vibes Technology, Molengraaffsingel 14, 2629 JD Delft, The Netherlands \\
	e-mail: \textbf{sklaassen@vibestechnology.com} \\
	$^3$ Gorenje d.o.o., Partizanska 12, 3503 Velenje, Slovenia \\
    e-mail: \textbf{tomaz.bregar@gorenje.com}}

\date{} 
\usepackage{geometry}
\usepackage{xcolor}

\usepackage[latin1]{inputenc}
\usepackage[T1]{fontenc}
\usepackage{array} 

\inputencoding{latin1} 

\usepackage{amsmath}
\usepackage{amsfonts}
\usepackage{amssymb}
\usepackage{bm} 
\usepackage{upgreek} 
\usepackage{units}
\usepackage{mathtools,leftidx}
\usepackage{longtable}
\usepackage{booktabs}
\usepackage{array,booktabs,enumitem}

\usepackage{customcommands}

\usepackage{graphicx}

\graphicspath{{figures/}{plots/}}     

\usepackage{cleveref}

\usepackage{subcaption}

\usepackage[bibencoding=utf8,url = false,  doi=false, isbn=false, style=numeric-comp]{biblatex}
\usepackage{float}

\begin{document}
\maketitle
\begin{abstract}
High quality measurements are paramount to a successful application of experimental techniques in structural dynamics. The presence of noise and disturbances can significantly distort the information stored in the data and, if not adequately treated, may result in erroneous findings and misleading predictions. A common technique to filter out noise relies on decomposing the dataset into singular components sorted by their degree of significance. Discarding low-value contributions helps to clean the data and remove spuriousness. This paper presents PRANK, a novel singular value-based reconstruction approach for multiple-response vibration datasets. PRANK integrates the effect of Principal Response Functions and Hankel filtering actions, resulting in an improved data reconstruction for both system poles and zeros. The proposed formulation is tested on both analytical and numerical examples, showcasing its robustness, efficiency and versatility. PRANK operates with both time- and frequency-based data. Applied to noisy full-field camera measurements, the filter delivered excellent performance, indicating its potential for various identification tasks and applications in vibration analysis.
\end{abstract}
\vspace{.4cm}
\textit{\textbf{Keywords}: PRANK, noise reduction, filtering, singular value decomposition, response functions}
\vspace{.4cm}

\vfill

\begingroup 
\setlength{\extrarowheight}{1pt}
\textbf{Nomenclature:} \newline \newline
\begin{small}
\begin{tabular}{l l l l}
	$\mat{Y}$ & time/frequency response matrix & MISO & multiple input single output \\
	$\mat{U}$,$\mat{V}$ & left, right singular vector matrix & MIMO & multiple input multiple output \\
	$\mat{\Sigma}$ & singular value matrix  & DoF & degree of Freedom	\\	
	$\mat{H}$ & Hankel realization  & FRF/IRF & frequency/impulse response function
	\\	
	$\mat{\sigma}$ & standard deviation  & TSVD & truncated singular value decomposition 	\\
	$[\star]_{i \times j}$  & matrix of size $i \times j$  & PRF & principal response function\\
	$(\star)_{ij}$  & entry $ij$ of the matrix   & SSA & singular spectrum analysis
	\\			
	SV & singular value & NVH & noise vibration and harshness  \\
	SVD & singular value decomposition & LSCF & least square complex frequency \\
	SIMO & single input multiple output & LSFD & least square frequency domain  \\
			
\end{tabular}
\end{small}
\endgroup

\newpage
\section{Introduction} 
\label{sec:Introduction}
The quality of measured data is paramount for the successful implementation of experimental-based techniques in the field of structural dynamics, vibration testing and modal analysis. The presence of measurement error can hinder a correct reading and interpretation of the findings by altering the underlying information stored in the acquired dataset. This, in turn, leads to major distortions when the data undergo laborious processing operations (e.g., matrix inversion). For this reason, techniques such as averaging and filtering are commonly adopted by experimentalists in the early stages of measurement processing \cite{Ewins2001, Allemang2020}. Among others, the use of the singular value decomposition (SVD) has gained popularity due to its effectiveness and ease of use. In essence, SVD factorizes a matrix in a way to understand the fundamental structure and relationships within the data by identifying the most critical features or directions along which the data varies the most \cite{Golub1996}. In this sense, the concept of SVD is closely related to the formulations of the Karhunen-Loeve decomposition \cite{Karhunen1947,Levy1948}, the principal component analysis \cite{Jolliffe1986} and the proper orthogonal decomposition \cite{Allen2018}, dominating statistics, data analysis and model reduction applications in various engineering fields. In the context of this article, the SVD enables the removal of low-valued contributions in terms of measured data reconstruction, which are usually associated with noise and spuriousity. This approach is known as truncated singular value decomposition (TSVD). Experimental dynamicists have been influenced by and benefited from the countless applications in such diverse fields as image processing, speech enhancement and recognition, and fluid dynamics \cite{Hermus2007, Rowayda2012, Brenden2015}. A popular approach in vibration analysis is to apply the TSVD on the spatial representation of a MIMO response dataset per frequency line. Applications of this strategy are numerous \cite{Otte1994, Ewins2001, Allemang2010}, although little evidence can be found in the literature of its reliable use for a pure noise and error removal purpose. For that reason, two different TSVD approaches particularly attracted the interest of the authors. The first strategy relies on the concept of principle response functions (PRFs) \cite{Pickrel1996,Ewins2001}. A singular value truncation on the PRF representation of a measured SIMO/MIMO response dataset has been investigated in \cite{Chakar2002,Allemang2010} and its benefits and pitfalls in terms of noise reduction have been highlighted. The second strategy is based on a rank-reduced Hankel representation of individual responses via singular value truncation. This method has been widely employed for time responses due to its remarkable success in eliminating noise of random nature \cite{Sanliturk2005,Markovsky2008,Gavin2013,Golafshan2016}. The process of selection of the dominant singular vectors/values is crucial for the success of the application. Among the numerous variants that appear in literature based on qualitative and/or quantitative indicators, a recently developed empirically-based algorithm, the e15 \cite{Epps2019,Epps2019a}, has attracted the authors' attention due to its automatability and scalability. \\ Although both a PRF-based filter and a Hankel-based filter have been used in the literature, they have not been fully exploited, according to the authors. In particular, the PRF filter was mainly applied to remove random noise. While the filtered result seems satisfactory, the cleaning action appears poor when compared with Hankel-based algorithms. Instead, the authors want to focus on the ability of PRF to recognize inconsistencies in multiple-response vibration datasets, making it a potential tool for poles/zeros reconstruction of error-polluted data. \\ The pursuit of a robust, widely applicable and adaptable filtering and reconstruction strategy for NVH MIMO measurements prompted the authors to propose PRANK. PRANK results from the fusion of \textbf{PR}F and H\textbf{ank}el techniques. This integration exploits their synergistic and complementary abilities to diminish random noise while detecting and removing outliers and distortions. A novel mixed approach, going beyond a sequential application of the filtering algorithms, is formulated to substantially reduce computational costs and improve versatility. The integration of an e15-based algorithm, which is proven by the authors to be effective for MIMO vibration datasets, resolves the inconveniences related to the truncation choice, thus delivering an automated filtering process. PRANK offers an non-physical alternative to classical modal filtering/reconstruction, the advantages and disadvantages of which are discussed in the article. \\ The idea of PRANK is given and a physical interpretation of the filtering action on dynamic measurements is proposed. The mathematical optimization underlying the TSVD process, the automatic selection of the reduction space, the different flavors and implementations of PRANK are discussed and subsequently tested on an analytical benchmark. The advantages and limitations of the proposed approach are addressed. A realistic MIMO numerical scenario is simulated to show the full capability of the algorithm. Finally, an experimental case highlights its usability in real world applications. \\
The filtering approach is mathematically described in \cref{sec:PRANK_sec}. An analytical proof-of-concept synthesizing FRFs is then presented in \cref{sec:SVR}. A numerical MIMO dataset explores the full capabilities of PRANK in \cref{sec:NE}. An experimental application of a full-field modal analysis with noisy high speed camera data is shown in \cref{sec:EE}. Discussion and conclusive remarks are collected in \cref{sec:Discussion} and \cref{sec:Conclusion}, respectively.

\section{PRANK: a robust, efficient and automated approach to reduce noise on measured response data via singular value truncation}
\label{sec:PRANK_sec}
The starting point is a SIMO/MISO/MIMO FRF, IRF or response dataset of the type $[\mathbf{Y}]_{n_o \times n_i \times n_k}$, where $n_o$, $n_i$ and $n_k$ are the number of outputs, inputs and spectral lines, respectively. Let's assume that the data are corrupted by measurement error and a filtering action preserving the underlying relevant dynamics is desired. The goal is on the accurate reconstruction of the system's poles and zeros (e.g., represented by peaks and antipeaks in a FRF), with emphasis on the latter, which are the most hindered by the presence of noise. This can be crucial in all applications involving subsequent processing of the data with inverse operations (e.g., experimental substructuring, transfer path analysis...). Moreover, let's assume that the use of polynomial fitting techniques, such as classic modal-based filtering, is limited by a high modal dense and damped system. \\ The idea of PRANK is to perform the data reconstruction solely based on truncated singular value decomposition algorithms. By expressing the data in different representation spaces, the obtained reconstruction is robust, smooth and suitable for both time or frequency domain. Efficiency and automation are two other aspects covered in this work that make PRANK a versatile and practical approach to tackling numerous NVH applications. \\
The key elements of PRANK are summarized as:
\begin{itemize}
\item Robustness. This is guaranteed by the use of both PRF and Hankel matrix representations, hence the name of the approach. Recognizing the synergy of filters when applied together is the idea behind PRANK (see \cref{sec:PRANK_1} and \cref{sec:PRANK_2}). 
\item Efficiency. Integrating the filtering actions into a single-step reconstruction process makes PRANK able to handle large datasets within reasonable computational effort. This opens the door to further enhancements of the filtering strategy (see \cref{sec:PRANK_2}).
\item Automation. The bottleneck of singular value based truncation techniques is an adequate and user-friendly selection strategy. The authors choose an approach recently developed in the context of particle image velocimetry in fluid dynamics, one that has demonstrated remarkable robustness in its implementation in combination with PRANK. This makes the use of PRANK extremely simple and automatable, as it requires minimal user input and expertise (see \cref{sec:PRANK_5}). 
\end{itemize}
In the following, the application of TSVD for $3$ different variants in the data topology is presented and a detailed comparison and interpretation between the techniques is offered. Then, the basic principle of PRANK is formulated and an efficient implementation of the filtering strategy is provided. 
Subsequently, a short review of singular value selection strategies is given and the chosen selection algorithm is briefly discussed. Some remarks on the implementation of PRANK conclude the chapter.

\subsection{Three flavors of TSVD}
\label{sec:PRANK_1}
A very common use of TSVD in structural vibration testing is to apply it to the input-output matrix per each frequency line of interest. This will be referred as 'classic' or frequency-based approach \cite{Otte1994, Ewins2001}. The starting representation space is therefore the following:
\begin{equation}
[\mathbf{Y}]_{n_o \times n_i}=
\begin{bmatrix}
\mathbf{Y}_{11} & \mathbf{Y}_{12} & \cdots & \mathbf{Y}_{1 n_i} \\    
\mathbf{Y}_{21} & \mathbf{Y}_{22} & \cdots & \mathbf{Y}_{2 n_i} \\  
\vdots & \vdots  &  & \vdots  \\ 
\mathbf{Y}_{n_o 1} & \mathbf{Y}_{n_o 2}  &  \cdots & \mathbf{Y}_{n_o n_i}   
\end{bmatrix}, \forall \omega \in n_k 
\end{equation}  
From here, a truncated SVD is applied on the 2D matrix (i.e., for a given frequency) and a total of $r$ singular components are retained. This operation can be summarized as:
\begin{equation}
\label{eq: classicTSVD}
\begin{array}{lll}
[\mathbf{Y}]^{\text{filt}}_{n_o \times n_i}&=&[\mathbf{U}]_{n_o \times n_r}[\mathbf{U}]^H_{n_o \times n_r}[\mathbf{Y}]_{n_o \times n_i}[\mathbf{V}]_{n_i \times n_r}[\mathbf{V}]^H_{n_i \times n_r}\\&=&[\mathbf{U}]_{n_o \times n_r}[\mathbf{\Sigma}]_{n_r \times n_r}[\mathbf{V}]^H_{n_i \times n_r}
\end{array}
\end{equation}
where $\mathbf{U}$, $\mathbf{\Sigma}$ and $\mathbf{V}$ are the left singular vector matrix, singular value matrix and right singular vector matrix, respectively, and where $n_r$ is the dimension of the truncated representation space. The column vectors in $\mathbf{U}$ and $\mathbf{V}$ can be interpreted as approximate mode shapes and modal participation factors. The associated singular values in $\mathbf{\Sigma}$ are scaling factors indicating the importance of the specific singular contribution. The superscript ${(\star)}^H$ denotes an Hermitian operation (i.e., conjugate transposition). The operation is repeated for each frequency line, and the amount of SV retained can theoretically vary by frequency as well.  \\
The filtering in \cref{eq: classicTSVD} can mathematically be interpreted as a two step process: A projection of the original dataset into some reduced realizations of the left and right singular vector matrices provides the reduced singular value matrix ($[\mathbf{\Sigma}]_{n_r \times n_r}=[\mathbf{U}]^H_{n_o \times n_r}[\mathbf{Y}]_{n_o \times n_i}[\mathbf{V}]_{n_i \times n_r})$. Then, an expansion with the same reduced realizations reconstructs the dataset in the original DoF space ($[\mathbf{U}]_{n_o \times n_r}[\mathbf{\Sigma}]_{n_r \times n_r}[\mathbf{V}]^H_{n_i \times n_r}$). The first step in this interpretation can be seen as the filtering process and consists in a left and right least-square realization, which filters out non-relevant dynamical contributions on a frequency basis. Due to the nature of the optimization problem, weighting matrices could potentially be added to act on the measured DoFs and give more importance to trusted data. Nonetheless, the presence of weighting matrices do not change the characteristics of the filtering process, which depends primarily on the representation space of the dataset. Hence, weighting will be ignored here and in all subsequent derivations of TSVD. \\  Overall, the filter can discriminate over frequency and is generally computationally cheap. Due to the ability of the technique to observe simultaneously the whole spatial dataset, it is beneficial in applications involving significant DoFs/data redundancy. However, it is too sharp for most common noise removal applications and, most importantly, is blind along the spectral axis, thus unable to discern and 'follow' physical modal information in the chosen frequency range. Indeed, a single modal contribution rises and falls in rank as it moves along the frequency axis, thus making this TSVD process generally spectrally discontinuous. \\

\begin{figure} 
	\centering
	\begin{subfigure}{\textwidth}
		\includegraphics[width=\textwidth]{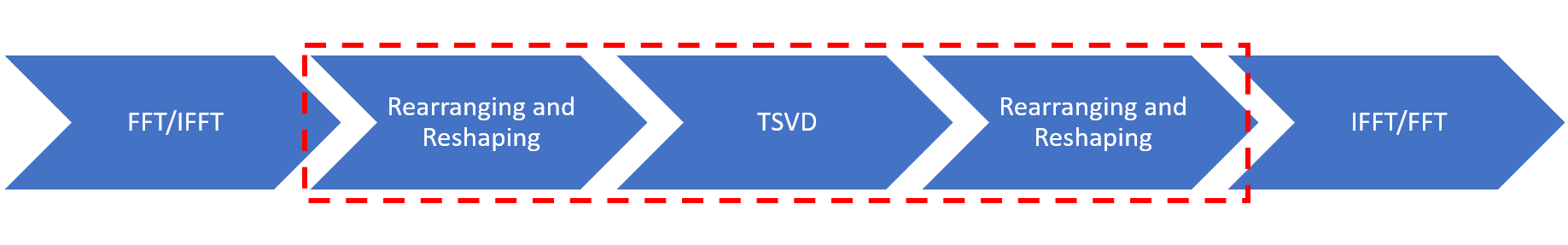}
	\end{subfigure}
	\caption{Filtering scheme for the PRF TSVD. FFT (Fast Fourier Transform) and IFFT (Inverse Fast Fourier Transform) for both input and output of the workflow represent the idea that PRF can be potentially applied on frequency- or time-based data.}
	\label{fig:scheme_PRF}
\end{figure}

A less common practice for the application of the TSVD relies on a transformation of the  representation space of the dataset by a simple rearranging and reshaping operation. Mathematically, the original dataset is 'flattened' as:
\begin{equation}
\label{eq:flattening}
[\mathbf{\bar{Y}}]_{n_k \times n_o n_i}=\left[{\mathbf{Y}_{11}(k)}|{\mathbf{Y}_{21}(k)}|\cdots|{\mathbf{Y}_{n_o n_i}(k) }     \right]
\end{equation}
The 3D matrix representation is transformed in a 2D representation, where the 'flattening' is applied on the spatial entries in the form of a column-based vectorization. Each column of the transformed dataset is a single FRF/IRF entry \footnote{Note that the PRF concept for noise reduction has only been used for FRF data in \cite{Allemang2010,Chakar2002}. Here, we expand the concept to IRF data without loss of accuracy, as will be shown in \cref{sec:SVR}.} ${\mathbf{Y}_{oi}(k)}$ unfolded along the spectral axis. Here, $k$ denotes the spectral index, which could be either time $t$ or frequency $\omega$.\\
The singular value truncation process is analogous to that of \cref{eq: classicTSVD} and writes:
\begin{equation}
\label{eq: PRF_TSVD}
\begin{array}{lll}
[\mathbf{\bar{Y}}]^{\text{filt}}_{n_k \times n_o n_i}&=&[\mathbf{U}]_{n_k \times n_r}[\mathbf{U}]^H_{n_k \times n_r}[\mathbf{\bar{Y}}]_{n_k \times n_o n_i}[\mathbf{V}]_{n_o n_i \times n_r}[\mathbf{V}]^H_{n_o n_i \times n_r}\\&=&[\mathbf{U}]_{n_k \times n_r}[\mathbf{\Sigma}]_{n_r \times n_r}[\mathbf{V}]^H_{n_o n_i \times n_r}
\end{array}
\end{equation}
where $r$ is the amount of retained singular component. Due to the topology transformation, the physical interpretation of the factorization terms is different from the one proposed by the classic approach. Here, the frequency distribution and the spatial distribution of the dynamical information are associated respectively to the left $\mathbf{U}$ and right $\mathbf{V}$ singular vector matrices. The corresponding singular values in $\mathbf{\Sigma}$ act again as an amplitude scale ranking by importance.\\ In this context, the Principal Response Functions (PRFs) are defined as the dominant singular vectors weighted by the associated singular values \cite{Ewins2001, Allemang2010}:
\begin{equation}
\label{eq:PRFs}
\text{PRF}_r=[\mathbf{U}]_{n_k \times n_r}[\mathbf{\Sigma}]_{n_r \times n_r}
\end{equation}
Once the filtering action is completed, the reconstruction and reorganization of the original 3D representation of the dataset follows. The filtering concept is called PRF TSVD and its implementation scheme is summarized in \cref{fig:scheme_PRF}. It is pointed out that both time and frequency response data could theoretically be fed to the algorithm.\\
The truncation on the PRFs preserves the primary dynamic characteristics of the full dataset and, at the same time, enables to exploit the spectral 'continuity' of the information. For that reason, this filtering tool is able to detect and remove distortions and outliers, as well as guaranteeing a moderate noise removal. A single SVD operation is performed. In comparison to the previous filter, the larger size of the flattened matrix undergoing SVD helps to smooth out the filtering action. This, in turn, is also the main limitation of the PRF TSVD, which often does not lead to a fully satisfactory noise reduction.\\

\begin{figure} 
	\centering
	\begin{subfigure}{1\textwidth}
		\centering
		\includegraphics[width=1\linewidth]{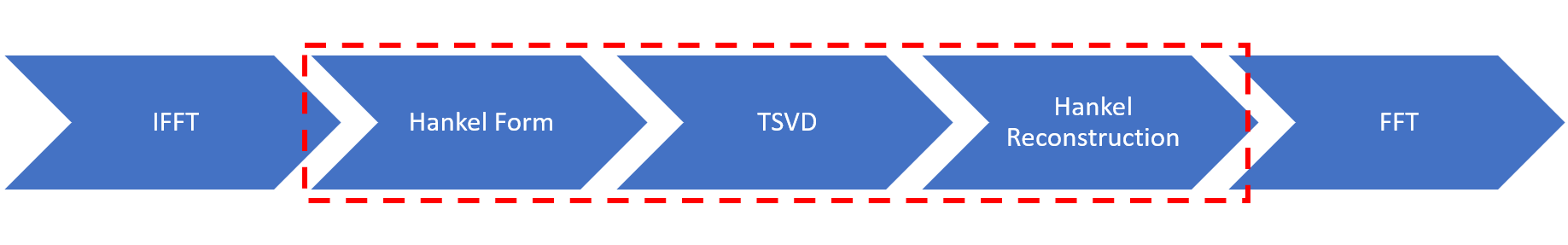}
	\end{subfigure}
	\caption{Filtering scheme for the Hankel TSVD.}
	\label{fig:scheme_Hankel}
\end{figure}

\begin{figure} 
	\centering
	\begin{subfigure}{0.6\textwidth}
		\centering
		\includegraphics[width=1\linewidth]{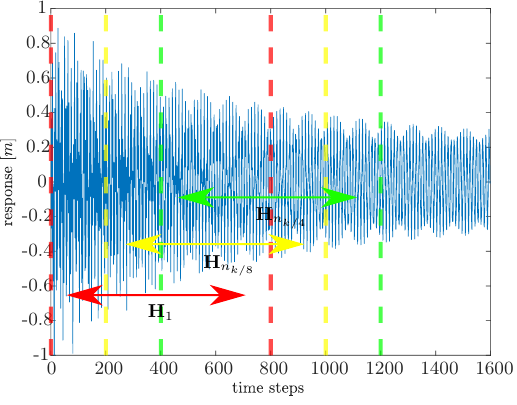}
	\end{subfigure}
	\caption{Example of snapshot subsets corresponding to column vectors of the Hankel matrix applied to an impulse response function.}
	\label{fig:Hankel_graphics}
\end{figure}

The third approach considered for the SVD filtering is based on the realization of a Hankel matrix, which consists of a matrix with the elements along each anti-diagonal equal. Taking a single spatial entry of the original dataset processed in the time domain, the corresponding Hankel form is constructed as follows: 
\begin{equation}
[\mathbf{H}]_{n_{k/2} \times n_{k/2}}= \left[{\mathbf{H}_{1}}|{\mathbf{H}_{2}}|\cdots|{\mathbf{H}_{n_{k/2}} }     \right]=
\begin{bmatrix}
Y_1 & Y_2 & \cdots & Y_{n_{k/2}} \\    
Y_2 & Y_3 & \cdots & Y_{n_{k/2+1}} \\  
\vdots & \vdots  &  & \vdots  \\ 
Y_{n_{k/2}} & Y_{n_{k/2+1}}  &  \cdots & Y_{n_k}   
\end{bmatrix}
\end{equation}
where $\mathbf{H}_{j}$ is a column vector of the square (and symmetric) Hankel matrix $[\mathbf{H}]_{n_{k/2} \times n_{k/2}}$ and $\mathbf{Y}_{oi}=\left[Y_1,Y_2,\cdots,Y_{n_k}\right]=\left[Y(t=t_1),Y(t=t_2),\cdots,Y(t=t_{n_k})\right]$ is an individual response function. Note that a Hankel realization can be extended to rectangular matrices \cite{Gavin2013}. \\
The SVD operation is then applied in analogy to \cref{eq: classicTSVD} and \cref{eq: PRF_TSVD} and reads:
\begin{equation}
\label{eq: Hankel_TSVD}
\begin{array}{lll}
[\mathbf{H}]^{\text{filt}}_{n_{k/2} \times n_{k/2}}&=&[\mathbf{U}]_{n_{k/2} \times n_r}[\mathbf{U}]^H_{n_{k/2} \times n_r}[\mathbf{H}]_{n_{k/2} \times n_{k/2}}[\mathbf{V}]_{n_{k/2} \times n_r}[\mathbf{V}]^H_{n_{k/2} \times n_r}\\&=&[\mathbf{U}]_{n_{k/2} \times n_r}[\mathbf{\Sigma}]_{n_r \times n_r}[\mathbf{V}]^H_{n_{k/2} \times n_r}
\end{array}
\end{equation}
After the filtering, the Hankel form is destroyed and must be reconstructed before rearranging the processed response to the original representation. The most common reconstruction strategy consists in an algebraic averaging along each anti-diagonal, also called Singular Spectrum Analysis (SSA) \cite{Golyandina2001,Markovsky2008}. Alternative processes could be employed like e.g. the Cadzow algorithm (repeated SSA) \cite{Cadzow1988} or a structured low-rank approximation (Total Least Square) \cite{LEMMERLING1999}. These latter will not be considered in this work. This filtering operation is called here Hankel TSVD and its implementation workflow is summarized in \cref{fig:scheme_Hankel}. The filter is particularly effective in removing random errors on individual response vectors while preserving the underlying dynamics. The principle of the cleaning process is quite basic: The available data is arranged in overlapping subsets of snapshots screening the entire response by single-step shifts. This is graphically illustrated in \cref{fig:Hankel_graphics}, where an impulse response is analyzed and $3$ overlapping snapshot subsets corresponding to column vector of the Hankel matrix are visualized. The decomposition process will extract the dynamic information consistently recurring throughout the batch of subsets and eliminate data of a random nature conversely. Therefore, this is suitable for temporal responses and less effective for frequency spectra. \\ Overall, the Hankel TSVD offers a smooth filtering of random noise. Nonetheless, its applicability is limited to single data entries, which makes it blind to the entire MIMO dataset (spatial discontinuity) and results in a computationally intensive procedure.

\subsection{PRANK as a mixed filtering strategy}
\label{sec:PRANK_2}
In the previous section, the $3$ flavours of TSVD were reviewed. Two of the three techniques appear to have complementary filtering actions and, according to the authors, their potential is not fully exploited in NVH applications. PRF TSVD is a simple but very effective tool for reconstructing signals based on a large set of spatially-distributed responses, while Hankel TSVD is well known as a powerful, albeit (in its presented form) spatially 'blind', random noise removal filter.\\\\
The most basic mathematical representation of PRANK involves applying both filters to the perturbed data sequentially. This process can be referred to as $PRANK_{HP}$ when the Hankel TSVD comes first followed by the PRF TSVD, or as $PRANK_{PH}$ when the PRF TSVD is applied first and then followed by the Hankel TSVD. Both the implementations, however, in their raw form suffer from the computational limitation associated with the Hankel part of the two-step approach. Indeed, if PRF requires the application of a single large SVD (of size $n_k \times n_o n_i$), Hankel needs a large SVD (of size $n_k \times n_k$ or $2 n_k \times 2 n_k$) to be applied individually on each entry of the spatial response dataset. Since $n_k$ is usually larger than or comparable to $n_o n_i$ for the applications considered by this article (with $n_k$ representing the time index), the computational burden is evident. A solution to this problem is presented in the following.\\
Consider the PRF TSVD process realized as flattening (\cref{eq:flattening}), reduced SVD and reconstruction (\cref{eq: PRF_TSVD}). Before performing the least square reconstruction, store the retained left singular deformation modes (or non-scaled PRFs, see \cref{eq:PRFs}) and apply a Hankel TSVD on each of them, individually. Take the cleaned left singular modes and finally reconstitute back the physical dataset in the original domain passing through the PRF reconstruction. This workflow is called here $PRANK_{HiP}$ (i.e., Hankel in PRF) and is summarized as:
\begin{enumerate}
\item Flattening the original dataset $[\mathbf{Y}]_{n_o \times n_i \times n_k}$ according to PRF as \begin{equation}
[\mathbf{\bar{Y}}]_{n_k \times n_o n_i}=\left[{\mathbf{Y}_{11}(k)}|{\mathbf{Y}_{21}(k)}|\cdots|{\mathbf{Y}_{n_o n_i}(k) }     \right]
\end{equation}.
\item Apply SVD on $[\mathbf{\bar{Y}}]$ and store the dominant $r$ left singular components $[\mathbf{U}]^{PRF}_{n_k \times n_r}$ (as well as the corresponding singular values $[\mathbf{\Sigma}]^{PRF}_{n_r \times n_r}$ and right singular vectors $[\mathbf{V}]^{PRF}_{n_o n_i \times n_r}$ ).
\item Apply a TSVD on each of the $r$ retained left singular components re-organized according to Hankel as \begin{equation}[\mathbf{H}]_{n_{k/2} \times n_{k/2}}=
\begin{bmatrix}
	\mathbf{u}_1 & \mathbf{u}_2 & \cdots & \mathbf{u}_{n_{k/2}} \\    
	\mathbf{u}_2 & \mathbf{u}_3 & \cdots & \mathbf{u}_{n_{k/2}+1} \\  
	\vdots & \vdots  &  & \vdots  \\ 
	\mathbf{u}_{n_{k/2}} & \mathbf{u}_{n_{k/2}+1}  &  \cdots & \mathbf{u}_{n_k}   
\end{bmatrix}\end{equation}  where $\mathbf{U}^{PRF}_{i}=\left[\mathbf{u}_1,\mathbf{u}_2,\cdots,\mathbf{u}_{n_k}\right]$ is the i-th individual PRF retained deformation mode. 
\item Reconstruct back the flattened representation from the $r$ cleaned left singular components $[\mathbf{U}]_{n_k \times n_r}^{Hankel}$ and the remaining PRF singular values/vectors as $[\mathbf{U}]_{n_k \times n_r}^{Hankel}[\mathbf{\Sigma}]_{n_r \times n_r}^{PRF}([\mathbf{V}]_{n_o n_i \times n_r}^{PRF})^H$. Finally, retrieve the cleaned dataset in the original domain as $[\mathbf{Y}]^{filt}_{n_o \times n_i \times n_k}$.
\end{enumerate}
Note that the spectral index $k$ in the above procedure might represent time or frequency. At the PRF stage, the difference is usually (at least for the cases considered in this article) negligible. Different is the case of the Hankel stage. In fact, in the previous section, the application of Hankel TSVD was limited to temporal responses only. The use of $PRANK_{HiP}$, on the other hand, allows a meaningful application of a frequency-based Hankel form. The difference is that, passing through the PRF vectors, 'individual' deformation modes are scanned by the Hankel-based process defined in frequency snapshot subsets. Hence, the dynamical behaviour is fully preserved. \\ 
This implementation drastically reduces the computational cost by decreasing the number of calls of the Hankel-based SVD from $n_o n_i$ to $n_r$. The benefits will generally increase as the number of spatial entries of the considered datasets increase. Indeed, it is likely that the number of significant dynamical contributions (in the defined spectral domain) represented by $n_r$ will tend to converge within a limited number of spatial measurement points (as soon as the limits of controllability and observability are overcome). Time can be easily cut by factors > $10\times$. An example will be given in \cref{sec:NE}. \\ Other efficiency options may be adopted to speed up the computation of the Hankel filtering stage: 
\begin{enumerate}
	\item Parallel processing: Using multiple cores in parallel to execute the individual Hankel-based SVDs.
	\item Use of 'thin/economy' SVD: Constructing and solving a single large (and rectangular) Hankel matrix by stacking each individual Hankel form column-wise. 
\end{enumerate}
While both solutions can reduce the computational burden of PRANK (in all its forms), they do so at the cost of increased memory requirements. Consequently, they will not be further discussed in this paper but will be integrated into future filter developments.\\\\

\subsection{Selection of dominant singular values and reconstruction}
\label{sec:PRANK_4}
The selection of singular components to retain and the reconstruction algorithm are key steps for a robust filtering. This section provides a brief review of the basic principle behind the common selection process and criteria.\\
All TSVD processes considered in \cref{sec:PRANK_1} (formulated in \cref{eq: classicTSVD}, \cref{eq: PRF_TSVD}, \cref{eq: Hankel_TSVD}) and further mixed in \cref{sec:PRANK_2} are based on a least square subspace-based lower rank realization of the input dataset \cite{Dendrinos1991, Hermus2007}. This corresponds to the best rank-$r$ approximation of the error-polluted dataset if and only if the following assumptions are met: the underlying clean signal is of order $r$, the noise is white and orthogonal to the signal. The last two assumptions, in particular, enable the decomposition of the processed dataset as:
\begin{equation}
\label{eq:LS_realization}
\mathbf{Y}=\mathbf{U}(\sqrt{\mathbf{\Sigma}^2_{cl}+\sigma_n^2 \mathbf{I}})\mathbf{V}^H
\end{equation}
where $\mathbf{\Sigma}^2_{cl}$ and $\sigma_n^2 \mathbf{I}$ are the matrices containing the squared singular values of the underlying clean signal and the noise variances, respectively. The simplification is in decomposing noise and signal with the same vector bases (white noise) and in the uncoupled addition of singular values of noise and signal (orthogonality). After the least-square based TSVD, the estimated dataset becomes $\mathbf{Y}^{\text{filt}}=\mathbf{U}_r \mathbf{\Sigma}_r \mathbf{V}_r^H$. The filtering is therefore based on a simple removal of the singular contributions $r+1..\text{end}$ (noise-only subspace), while maintaining unaltered the remaining $1..r$ singular values and corresponding singular vectors (signal + noise subspace). This work does not explore the sensitivity to the failure to meet the mentioned conditions. Note that alternative approaches to the simple least square estimation that further reduce the noise contribution in the remaining singular subspace, as well as extensions to properly treat coloured noise exist in the literature \cite{Hermus2007,VanHuffel1993,DeMoor1993,Hansen1999,Lev-Ari2003}. \\
The other thing that remains to be discussed is how to select the singular components to be retained. In principle, both qualitative and quantitative approaches may be adopted. For the former, the most common criterion is based on detecting the 'elbow' or turning point of the singular value curve (if existing). This is also known as Scree plot criterion and corresponds to a visual threshold for the truncation. A valid alternative specific for the PRF TSVD is the evaluation of the dominant PRFs (left singular vector in the corresponding PRF representation). Indeed, single PRFs can be viewed equivalently as principal spectral responses of single DoF oscillators and their degree of cleanliness/smoothness can easily be assessed visually. Examples of qualitative screening will be presented in \cref{subsec:SVR_1}. \\ On the other hand, quantitative criteria may offer greater robustness and automation capabilities. Those can be classified in error-norm and error-free based approaches. The former requires the knowledge of an error threshold a priori, which could act as a representative noise floor. Values may be obtained via statistical (variance, standard deviation) or consistency (coherence value) quantities. An important remark is that it is necessary to derive the error norm in the appropriate representation of the dataset prior to filtering (classic, PRF or Hankel). Opposite to that, the error-free criteria have no prior knowledge of the error/noise function and determine the removal threshold based on the relative importance of the singular contributions to reconstruct the full information. Quantitative criteria compare the determined (or chosen) removal threshold with the singular values magnitude. As an alternative, the singular values cumulative magnitude (sum of the last $n-r$ singular values) may be adopted. Examples of quantitative estimations will be shown in \cref{subsec:SVR_1}. 

\subsection{Automating the selection process in PRANK}
\label{sec:PRANK_5}
To speed-up and automate the use of PRANK, the authors propose to implement a selection and reconstruction algorithm recently developed for the treatment of large batches of data in fluid dynamics applications. The so-called e15 algorithm \cite{Epps2019,Epps2019a} is an empirical procedure that embeds a self-recognition of a noise floor from the perturbed data, an estimation of the error contribution for each singular component, and finally a selection/reconstruction of the dominant ones. \\ The procedure can be summarized as:
\begin{enumerate}
\item Perform an SVD on the noisy data.
\item Estimate the measurement error (noise floor) and the spatial correlation parameter (correlation index) by fitting the Marchenko-Pastur distribution \cite{Marcenko1967} to the tail of the noisy singular values.
\item Estimate the root mean square error of each noisy singular mode given the parameters estimated in $2.$ and the batch size.
\item Estimate the rank $r$ for minimum-loss reconstruction using the root mean square estimate in $3.$ given the error ceiling and a user pre-defined threshold parameter (tolerance $\mu$)
\item Reconstruct an estimate of the first $r$ retained clean singular values by removing the fitted Marchenko-Pastur singular values from the noisy singular values.
\item Reconstruct back the clean data in the original domain.
\end{enumerate}
For a deeper understanding of the above-mentioned points, the reader is referred to \cite{Epps2019,Epps2019a}.\\
Recalling the classification of techniques depicted in \cref{sec:PRANK_4}, e15 is a quantitative error-free based approach. Indeed, no a priori knowledge is needed to formulate a noise floor/ceiling. The latter is directly extracted from the noisy data, arranged in the form of its singular value curve, by a best-fit of a statistical empirical distribution, i.e., the Marchenko-Pastur \cite{Marcenko1967}. This information is then used to determine the cleanliness of the noisy singular vectors (according to the formulation derived in \cite{Epps2019}), which are assigned an index from $0$, for the noise ceiling, to $1$, for the first dominant/cleanest mode. At this stage, a user-defined parameter, namely a threshold value, is used to determine the modes to be retained. This can be considered the only piece of information that requires (user) attention, although it has been shown that a high success rate can be achieved by employing some default values \cite{Epps2019a}. Finally, a SVD reconstruction is performed. Unlike a minimal least-squares realization, such as the one discussed in \cref{sec:PRANK_4}, the noise contribution is also subtracted from the retained singular subspace. This translates in a further cleaning of the dominant singular values, while singular vectors are kept unaltered.\\
To the authors' knowledge, the application of e15 has been limited to fluid dynamic applications and, in particular, has been developed and tested on particle image velocimetry datasets. Its reliability or robustness in handling FRF/IRF dynamic datasets isn't guaranteed. The backbone of the selection strategy is the Marchenko-Pastur distribution. This, by definition, empirically represents the noisy tail of the singular value curve when dealing with independent identically distributed random data. In PRANK, the spatial and/or spectral randomness may not comply with the above-mentioned assumption due to the re-organization of the 3D matrix onto a PRF and a Hankel form. Although the developers of e15 provide a correlation parameter that may help to modify accordingly the shape of the distribution, a better empirical fit of the statistical distribution may be found to adapt typical NVH datasets to the specific PRANK representation domains. This aspect, however, has not been pursued by the authors of this article, due to the satisfactory performance of e15 when applied within the filtering framework. \\ Examples of the application of e15 with PRANK will be shown in \cref{sec:NE}.

\subsection{Remarks about the implementation}
\label{sec:PRANK_6}
PRANK has been implemented as a filtering function for 3D IRF/FRF response datasets in Python and Matlab. The Python version is made available in the open-source toolbox pyFBS \cite{Bregar2022}. The following features are provided:
\begin{itemize}
\item Type of filtering action: Sequential application ($PRANK_{PH}$ or $PRANK_{HP}$) or mixed formulation ($PRANK_{HiP}$).
\item Spectral domain: time- or frequency-based application.
\item Type of selection strategy: Qualitative-based Scree plot evaluation or advanced automatic e15 evaluation.
\end{itemize}
The implementation of the e15 follows the derivation in \cite{Epps2019,Epps2019a}. The spatial correlation parameter loop has been adjusted on an empirical basis after the evaluation of multiple test scenarios with the type of datasets discussed within the article. The threshold parameter for the removal of singular components in the e15, which remains the only user-defined input in the process, is recommended to be kept between 5 to 10 percent.\\
The SVD algorithm chosen and the number of SVD calls depend on the above-mentioned options. The authors envision the use of a mixed time-based formulation with e15 as a default design.
\section{PRANK on a multi-DoFs analytical system}
\label{sec:SVR}
\begin{figure} [h]
	\centering
	\begin{subfigure}{0.8\textwidth}
		\centering
		\includegraphics[width=1\linewidth]{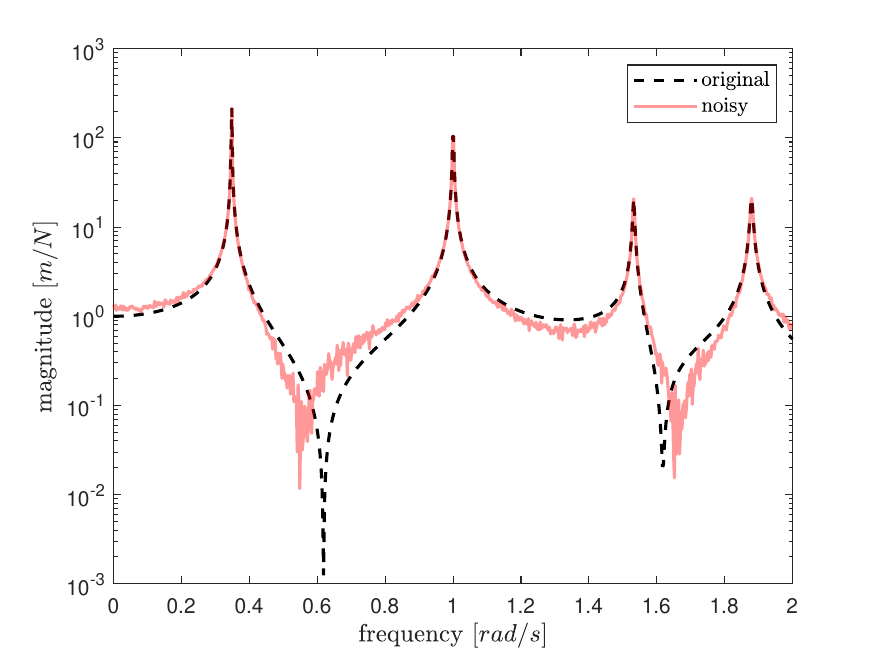}
	\end{subfigure}
	\caption{Comparison magnitude original and noisy FRF $\mathbf{Y}_{21}$.}
	\label{fig:noise}
\end{figure}

A simple analytical example is provided to explore some of the theoretical concepts discussed throughout \cref{sec:PRANK_sec}. The automation aspects will be treated in \cref{sec:NE}. \\ 
Frequency response functions are synthesized from a lightly-damped 4-DoF mass-spring-damper system. The parameters are listed in \cref{tab:parameters}. A complex-valued Gaussian random distributed noise is added to each FRF frequency by frequency. The corresponding standard deviations have the following form \cite{Trainotti2020}:
\begin{equation}
\label{eq:error}
\begin{array}{lcc}
\sigma_{ij,\Re}=a|FRF_{ij}|+b \\
\sigma_{ij,\Im}=c|FRF_{ij}|+d
\end{array}
\end{equation}
where $\sigma_{ij,\Re}$, $\sigma_{ij,\Im}$ are the error on the real and imaginary part and $a=0.003,b=0.06,c=0.003,d=0.05$.\\
This random error represents the noise on input and output of the acquisition chain and is assumed to be uncorrelated for simplicity. In addition, faulty and distorted responses are simulated by adding a real offset on top of all the FRFs with output DoF $2$ ($0.22$, $0.16$, $0.18$, $0.16$ on top of $\mathbf{Y}_{21},\mathbf{Y}_{22},\mathbf{Y}_{23},\mathbf{Y}_{24}$). Overall, the choice of the error function aims to highlight the capabilities of the different stages of the filter. Hence, both a pure random global contribution and local outliers/distortions are considered.\\
A comparison between the original and noise-polluted biased data is shown for an example FRF in \cref{fig:noise}. Both the effect of noise, which cuts the zeros and adds uncertainty to the overall curve, and the effect of the outlier, which shifts the zeros along the frequency axis, are evident. From now on, the distorted FRF will be referred as 'noisy' for simplicity.

\setlength{\extrarowheight}{2pt}
\setlength{\tabcolsep}{6pt}
\newcolumntype{P}[1]{>{\centering\arraybackslash}p{#1}}

\begin{table} [h]
	\centering
	\caption{\small{Parameters of the 4-DoFs analytical system.} }
	\begin{tabular}{ p{2cm}P{1cm}P{1cm}P{1cm} }
		\hline 
		DoFs  &  m &  d &  k \\
		\hline
		1 & 1 & 0.002 & 1\\
		2 & 1 & 0.002 & 1\\
		3 & 1  &  0.002 & 1 \\
		4 & 1 & 0.002 & 1\\
		\hline 
	\end{tabular}
	\label{tab:parameters}
\end{table}

\begin{figure} [h]
	\centering
	\begin{subfigure}{0.6\textwidth}
		\centering
		\includegraphics[width=1\linewidth]{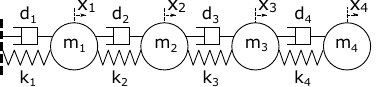}
	\end{subfigure}
	\caption{Analytical 4 DoFs mass-spring-damper system fixed at one end.}
	\label{fig:analytical}
\end{figure}

\subsection{Comparison between the different flavours of TSVD and PRANK}
\label{subsec:SVR_1}
\begin{figure}  [h]
	\centering
	\begin{subfigure}{0.495\textwidth}
		\centering
		\includegraphics[width=1\linewidth]{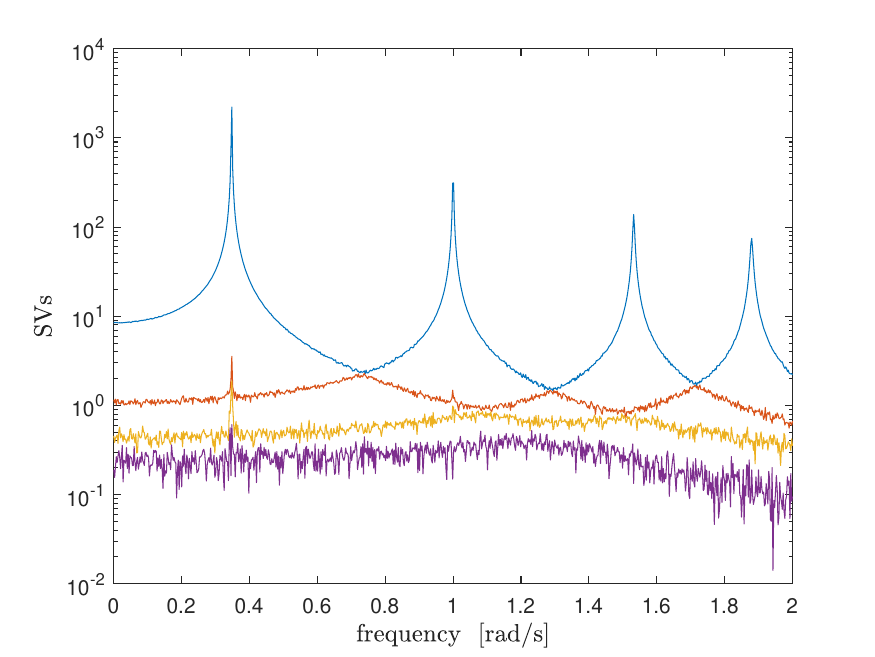}
	\end{subfigure}
	\begin{subfigure}{0.495\textwidth}
		\centering
		\includegraphics[width=1\linewidth]{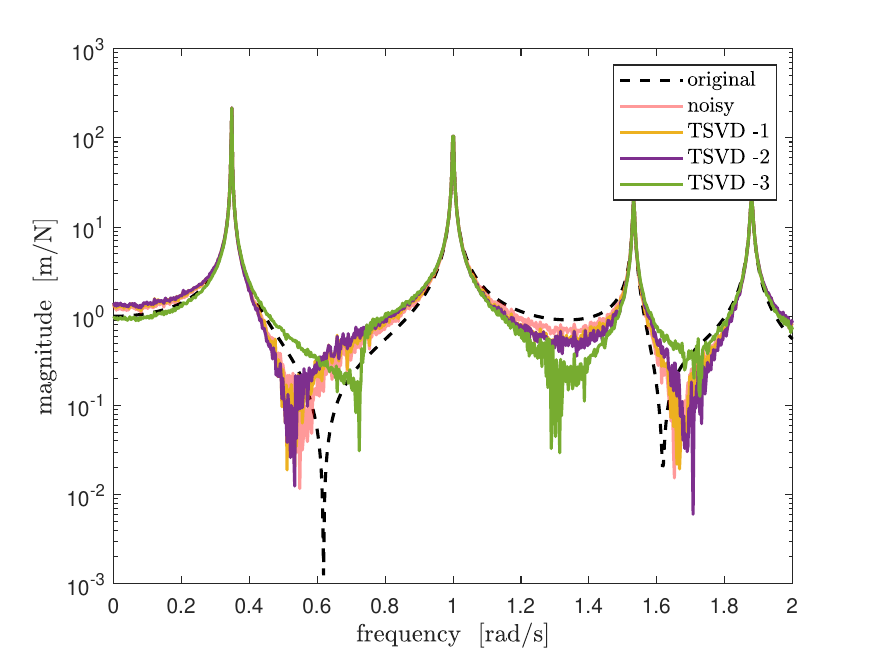}
	\end{subfigure}
	\caption{Effect of filtering with a classic TSVD filtering. Left: CMIF plot for the whole FRF dataset in the frequency range of interest. Right: The example FRF $\mathbf{Y}_{21}$ is plotted. A constant amount of SVs is removed for the whole frequency range for simplicity. The progressive removal of low-value SVs is shown in yellow (-1 SVs), pink (-2 SVs) and green (-3 SVs).}
	\label{fig:classic TSVD}
\end{figure}

The classic frequency-by-frequency use of the TSVD is analyzed. The plot of the singular values per frequency line (also called Complex Mode Indicator Function, or CMIF) is depicted in \cref{fig:classic TSVD} (left plot). Retaining $1$,$2$ or $3$ singular components per frequency line does not lead to significant differences in terms of noise and/or outliers removal, as shown in \cref{fig:classic TSVD} (right plot). The failure of the classic approach on the analytical example could be explained by an actual lack of redundancy in the dataset. Indeed, it could be hypothesized that all $4$ modes (hence a rank $4$ information) are contributing to build the response at every frequency line in the range of interest. Choosing a different amount of retained singular vectors per frequency interval is a tedious process and would not be of much value here. TSVD users for frequency response datasets often apply the technique before an inversion step, when there is a large redundancy of dynamic information in the data \cite{Otte1991,Otte1994}. This operation is interpreted as a regularization and it is beyond the scope of this paper.

\begin{figure}  [h]
	\centering
	\begin{subfigure}{0.495\textwidth}
		\centering
		\includegraphics[width=1\linewidth]{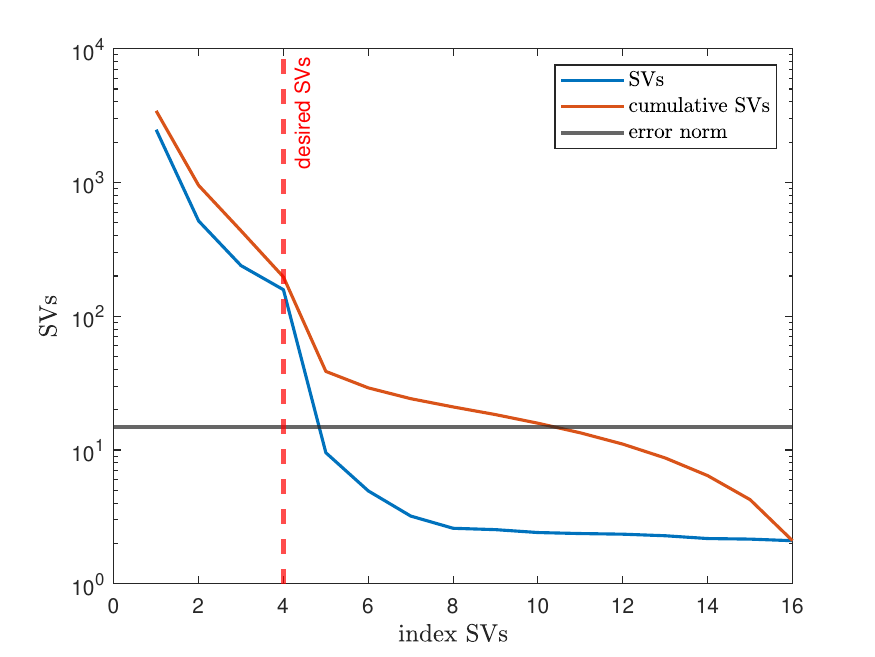}
	\end{subfigure}
	\begin{subfigure}{0.495\textwidth}
		\centering
		\includegraphics[width=1\linewidth]{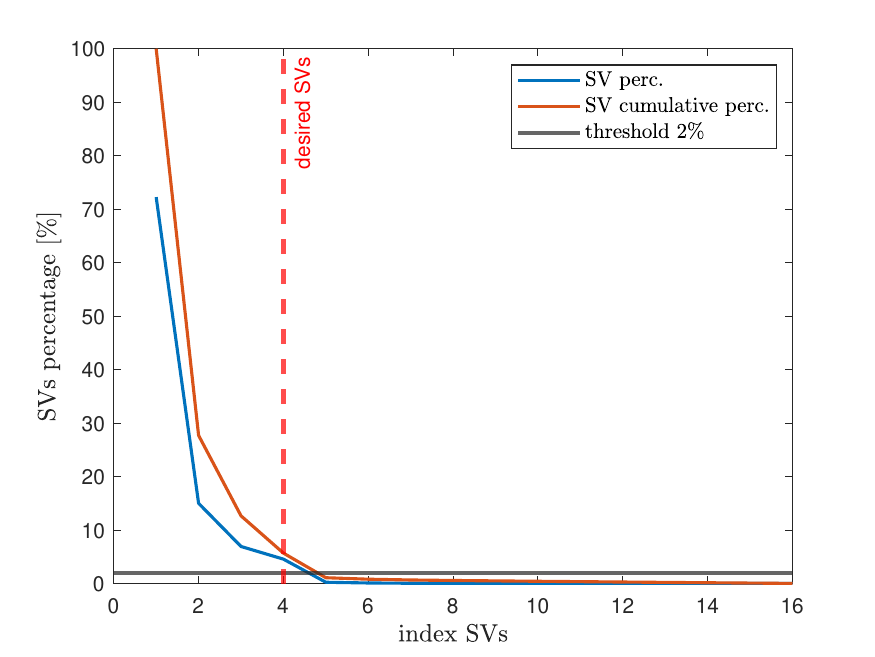}
	\end{subfigure}
	\caption{Quantitative selection of dominant singular values. Example on PRF representation. Left: Error-norm based approach. Selection based on the SV curve (blue) or the cumulative SV curve (orange). Right: Error-free based approach. Selection based on the SV relative curve (blue) or the cumulative SV relative curve (orange).}
	\label{fig:SV selection quantitative}
\end{figure}

If the classic TSVD fails in approaching this type of dataset, PRANK offers an alternative solution. Before showing the results of applying the proposed filter, the focus is on the selection of the singular components to retain within each TSVD operation. The selection phase was conceptually discussed in \cref{sec:PRANK_4}. Quantitative and qualitative approaches may be employed. In the following, some practical examples are provided using the analytical test-case.\\
Two quantitative techniques are compared in \cref{fig:SV selection quantitative}. The example is based on a PRF TSVD filtering. Similar considerations apply to the Hankel-based procedure. \\ On the left plot, an error-norm based approach is proposed. The threshold in the example is found by taking the error function/matrix used to perturb the original data and computing the norm of the 'flattened' representation according to the PRF strategy. 
Both the singular value curve (blue) and the cumulative singular value curve (orange) are plotted. Among the $2$ curves, the blue one seems to clearly suggest preserving only the first $4$ SVs (above the error threshold in black). This is in line with the expectations, since only $4$ modes should be relevant for re-constructing the dynamical info along the spectral axis.\\ On the right plot, an error-free based approach is provided. Since no absolute error norm is available, the threshold for removal is based on the relative importance of a certain singular value (blue) or cumulative singular value (orange). With a conservative value of $2\%$ as threshold, both curves find the desired number of relevant SVs.

\begin{figure}  [H]
	\centering
	\begin{subfigure}{0.495\textwidth}
		\centering
		\includegraphics[width=1\linewidth]{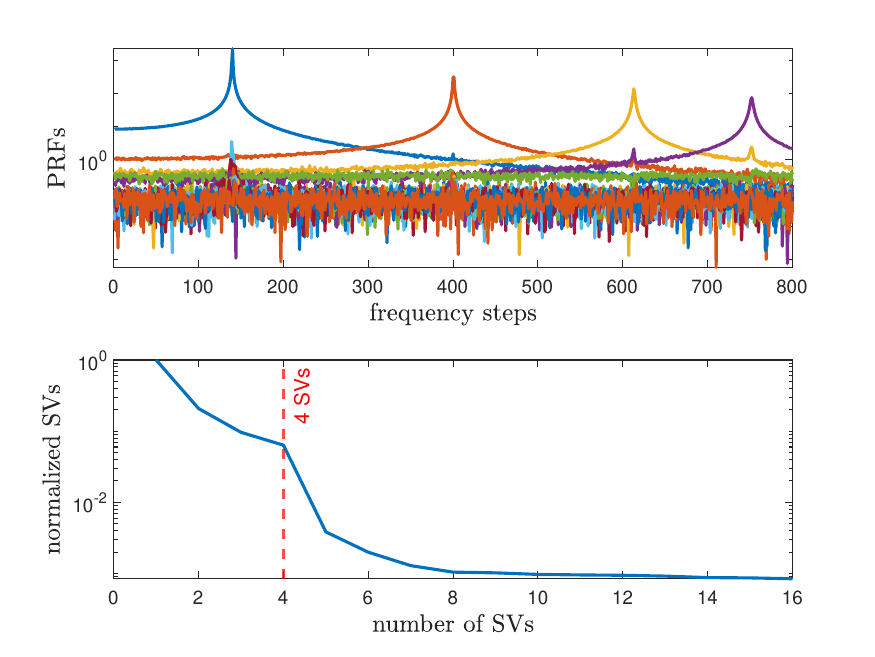}
	\end{subfigure}
	\begin{subfigure}{0.495\textwidth}
		\centering
		\includegraphics[width=1\linewidth]{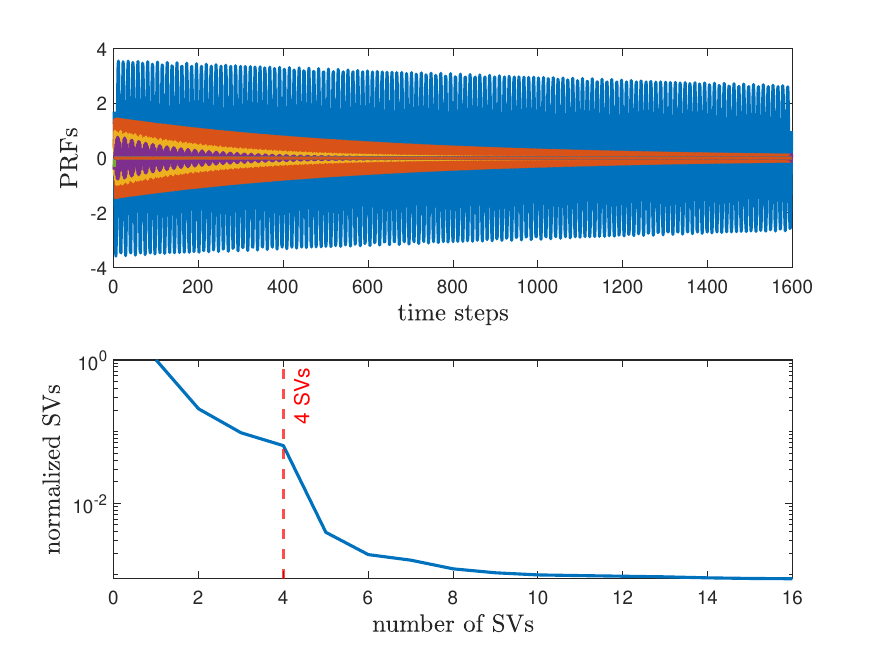}
	\end{subfigure}
	\caption{Qualitative selection of dominant singular values for PRF TSVD. Example on noisy FRF dataset. Left: frequency-based filtering. Principal Response curves on top, singular value curve with chosen SVs on bottom. Right: time-based filtering. Principal Response curves on top, singular value curve with chosen SVs on bottom.}
	\label{fig:PRF_selection}
\end{figure}

\begin{figure} [H] 
	\centering
	\begin{subfigure}{0.495\textwidth}
		\centering
		\includegraphics[width=1\linewidth]{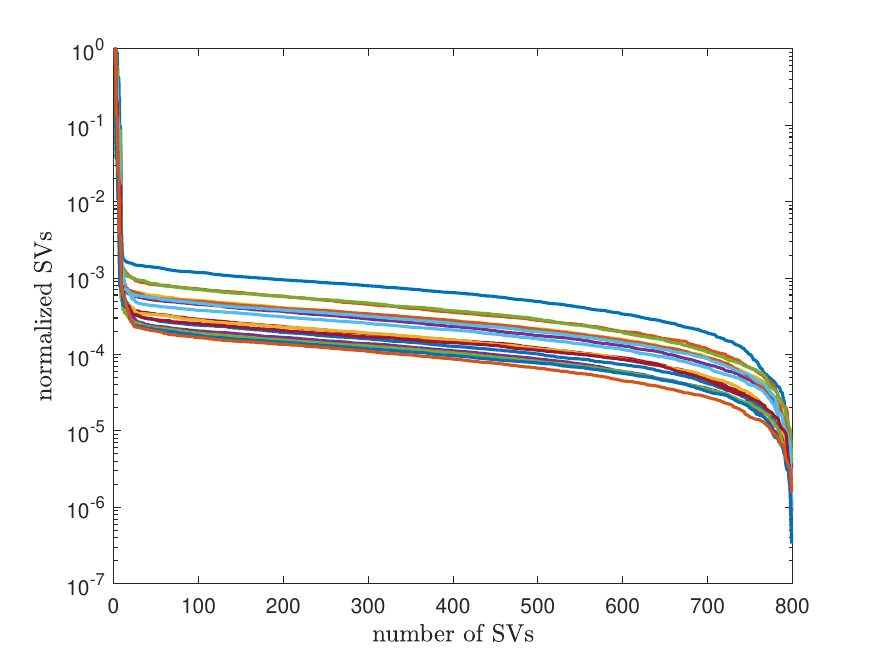}
	\end{subfigure}
	\begin{subfigure}{0.495\textwidth}
		\centering
		\includegraphics[width=1\linewidth]{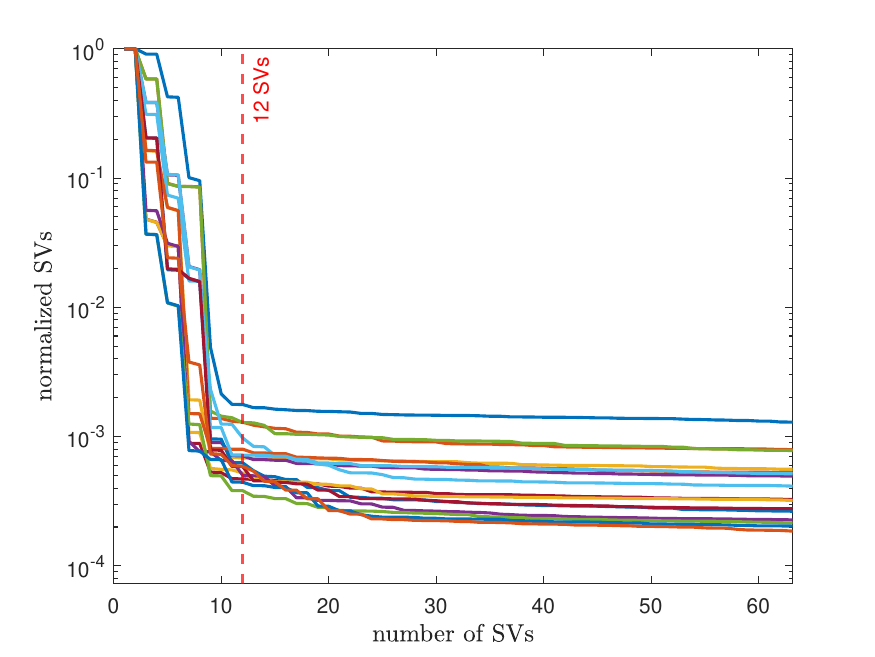}
	\end{subfigure}
	\caption{Qualitative selection of dominant singular values for Hankel TSVD with time-based filtering. Example on noisy IRF dataset. Left: Singular value curves for each IRF. Right: Magnification on chosen SVs.}
\label{fig:Hankel_selection}
\end{figure}

\begin{figure} [H] 
	\centering
	\begin{subfigure}{0.8\textwidth}
		\centering
		\includegraphics[width=1\linewidth]{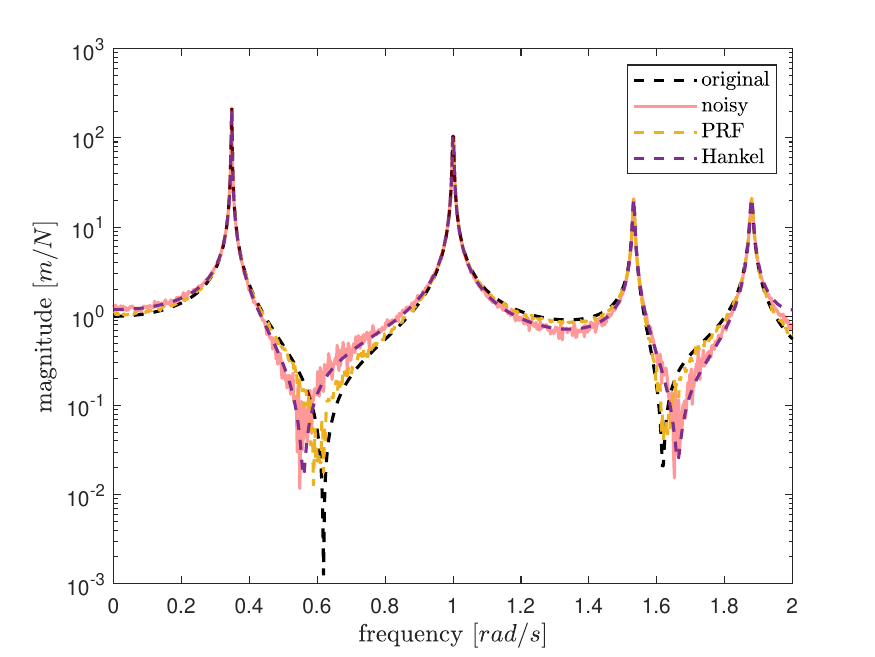}
	\end{subfigure}
	\caption{Comparison magnitude of FRF $\mathbf{Y}_{21}$ for original (black), noisy (pink), PRF-only (yellow), Hankel-only (purple) data.}
	\label{fig:comparison_21_1}
\end{figure}

\begin{figure} [H] 
	\centering
	\begin{subfigure}{0.8\textwidth}
		\centering
		\includegraphics[width=1\linewidth]{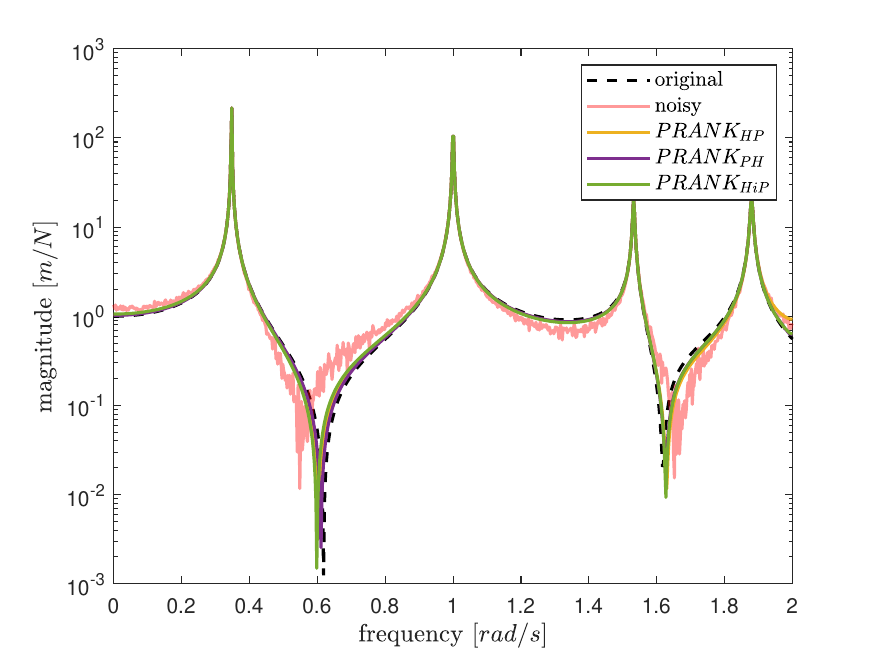}
	\end{subfigure}
	\caption{Comparison magnitude of FRF $\mathbf{Y}_{21}$ for original (black), noisy (pink), $PRANK_{HP}$ (yellow), $PRANK_{PH}$ (purple) and $PRANK_{HiP}$ (green) data.}
	\label{fig:comparison_21_2}
\end{figure}

A qualitative approach for the selection of SVs within the PRF TSVD is provided in \cref{fig:PRF_selection}. Both frequency-based (left plot, using FRFs) and time-based (right plot, using IRFs) filtering are shown. Besides the traditional Scree plot, the principal response function visualization could provide a more impactful insight into the pertinent spectrally-distributed dynamics. For a non-trained eye, looking at frequency-based PRFs ('single-DoF frequency oscillators') may be easier than time-based PRFs ('single-DoF time oscillators'). In the analyzed case, the choice of $4$ dominant components is rather revealing.\\
An analogous qualitative screening of singular values is commonly used for the Hankel TSVD. A reasonable assumption is to consider the noise floor as uniform over the full spatial dataset. This simplifies the selection process, as it avoids the need to evaluate each IRF separately. This is often valid, as reflected in \cref{fig:Hankel_selection}, where the 'nose' in the singular value curve seems to approximately occur for the same amount of SVs (in this specific case $12$ SVs is considered a conservative choice) for all responses.\\ Similar considerations are obtained by visually screening the singular value curves when applying the PRANK filtering options.\\\\ 
Finally, the results of filtering are compared in \cref{fig:comparison_21_1} and \cref{fig:comparison_21_2} with respect to the reference (black) and noisy (pink) FRF magnitude. The number of SVs is chosen to be $4$ for the PRF step and $12$ for the Hankel step (regardless of the filtering applied), based on the points previously discussed. The PRF-only (yellow) and Hankel-only (purple) filtering (\cref{fig:comparison_21_1}) follows expectations. The former is able to correct the location of zeros, although the attenuation of noise is not too satisfactory. The opposite is true for the Hankel filter. \\
It is clear from \cref{fig:comparison_21_2} that combining the action of the two filters in PRANK leads to a higher quality result. All variants proposed, namely $PRANK_{HP}$ (yellow), $PRANK_{PH}$ (purple) and $PRANK_{HiP}$ (green), reveal a similar accuracy. The PRF filter is able to detect and remove the faultiness (in this case represented by a wrong zero location) thanks to its global 'view' on the dynamical information of the full dataset. The Hankel filter removes the residual random error, thus enabling a clean reconstruction of the FRF. Similar observations are found for all other FRF curves. 

T
A visual/qualitative SVD selection and reconstruction strategy is a straightforward choice for the simple analytical benchmark under investigation. In a more realistic scenario, a robust and automated process like the one presented in \cref{sec:PRANK_5} is recommended. In addition, it is highlighted that, whenever an actual systematic error is present in the dataset, PRF (and consequently PRANK) would not be able to recognize any 'flaw' in the overall dynamics.  

\subsection{SVD-based data cleaning and physicality preservation}
\label{subsec:SVR_2}
\begin{figure} [H] 
	\centering
	\begin{subfigure}{0.8\textwidth}
		\centering
		\includegraphics[width=1\linewidth]{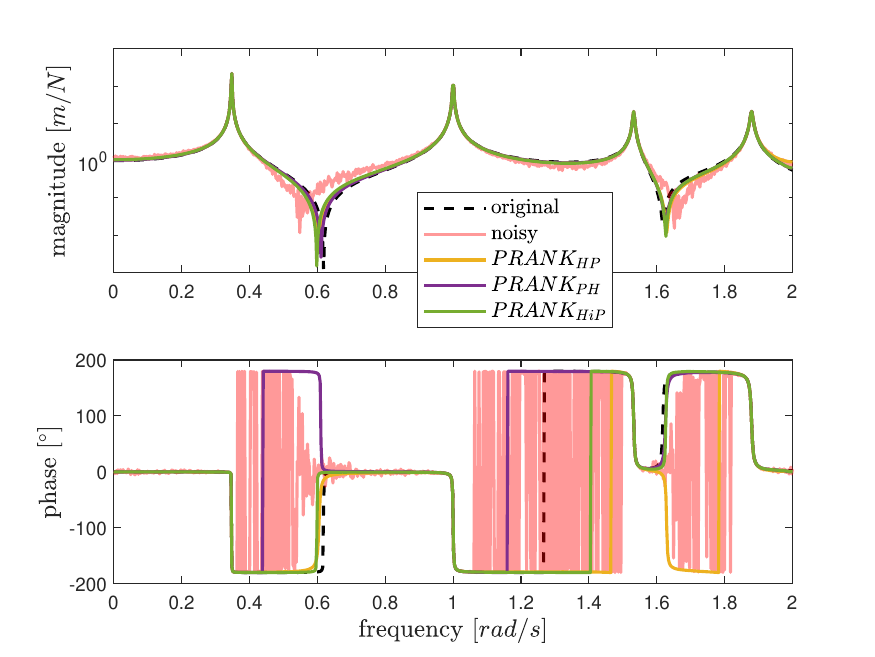}
	\end{subfigure}
	\caption{Comparison magnitude and phase of FRF $\mathbf{Y}_{21}$ for original (black), noisy (pink), $PRANK_{HP}$ (yellow), $PRANK_{PH}$ (purple) and $PRANK_{HiP}$ (green) data.}
\label{fig:comparison_21_absphase}
\end{figure}

\begin{figure} [H] 
	\centering
	\begin{subfigure}{0.7\textwidth}
		\centering
		\includegraphics[width=1\linewidth]{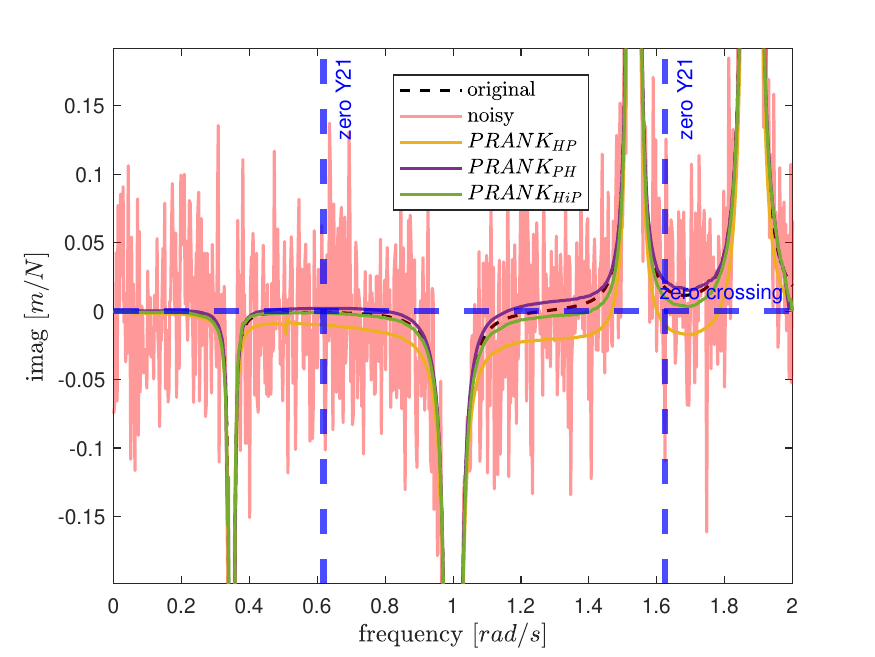}
	\end{subfigure}
	\caption{Comparison imaginary part of FRF $\mathbf{Y}_{21}$ for original (black), noisy (pink), $PRANK_{HP}$ (yellow) and $PRANK_{PH}$ (purple), $PRANK_{HiP}$ (green) data. Zoom around the X-axis zero crossing line and the zeros of the FRFs.}
\label{fig:comparison_21_realimag}
\end{figure}

The PRANK filtering has a limiting factor, which is common to any other SVD-based processing technique. The data after the TSVD treatment need not comply with physical conditions of linear time-invariant systems. In other words, no enforcement of reciprocity and passivity is ensured by the singular value decomposition. This will be further discussed in \cref{sec:Discussion} in comparison to modal-based decomposition techniques. \\
To illustrate this limitation, the phase of the filtered FRF of \cref{fig:comparison_21_2} is shown together with the magnitude in \cref{fig:comparison_21_absphase}. In proximity of the zeros of the FRF, the filtered responses may take a wrong turn (180 degree shift in the opposite direction) on a seemingly random basis (see the example of the purple curve close to the zero at ca. \unit[0.6]{rad/s} or the yellow curve close to the zero at ca. \unit[1.6]{rad/s}). The source of the issue is easily found in the randomness of the noisy data where the signal-to-noise ratio is extremely low (i.e. the zeros of the function). A visual explanation is offered in \cref{fig:comparison_21_realimag}, where the imaginary part of the FRF is magnified around the x-axis zero crossing line. Around the zero at ca. \unit[0.6]{rad/s} (green vertical line), the purple curve crosses the zero-line (green horizontal line). Similar is the case for the yellow curve around the zero at ca. \unit[1.6]{rad/s}. Being opposite in sign from the reference (black line) exactly at the zero of the FRF leads to a 180 degree turn in the opposite direction in phase. This erroneous behavior is driven by two factors. First, the signal-to-noise ratio is significantly lower than $1$. Second, the imaginary part of the FRF around a dynamical zero is very close to the zero-line (i.e. small damping condition). Phase errors when both conditions are met are therefore almost inevitable and random in nature. The authors verified that by increasing the damping of the system, these crossovers do not occur and the phases after filtering are reconstructed correctly.\\ 
The issue described by this example comes in the form of a lack of passivity. This may be critical when filtered data undergo processing steps involving inversions, as zeros become relevant. An example would be the use of PRANK within frequency-based substructuring. This is beyond the scope of this article and will not be addressed here.

%
\section{PRANK on a numerical example with a realistic design of experiments}
\label{sec:NE}
A numerical example is set up to test the usability of PRANK in a more realistic scenario. PRANK is evaluated as a filtering function applied on a noisy 3D response dataset in a structural testing environment. Focus is on robustness, efficiency and automation capabilities. The whole analysis inclusive of plotting and evaluation environments is performed using pyFBS \cite{Bregar2022}.

\subsection{Numerical benchmark and design of experiments}
\label{subsec:NE_1}
\begin{figure} 
	\centering
	\begin{subfigure}{0.495\textwidth}
		\centering
		\includegraphics[width=1\linewidth]{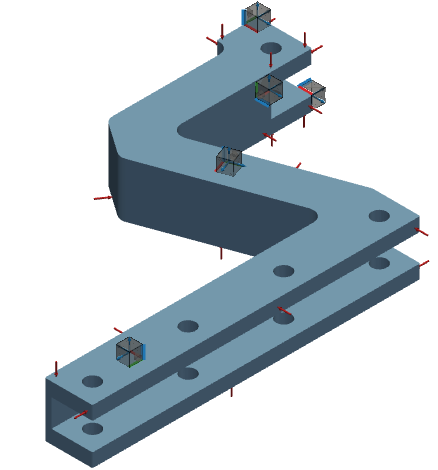}
	\end{subfigure}
	\caption{Benchmark structure with outputs (triaxial sensor box) and inputs (red arrow).} 
	\label{fig:CAD}
\end{figure}

The structure consists of a 3D extruded aluminum profile of a rectangular cross-section and is depicted in \cref{fig:CAD}. The system is modeled with finite elements and analyzed in free-free conditions. The measurement setup is simulated in pyFBS and consists of $21$ output ($7$ triaxial sensors in \cref{fig:CAD}) and $23$ input DoFs. Among them, a total of $9$ output and $10$ input DoFs are distributed throughout the whole surface of the structure, while the rest of the DoFs is clustered around one end.\\ 
The FRFs are synthesized from \unit[0-2000]{Hz} with \unit[1]{Hz} resolution via mode superposition using the first $100$ modes calculated with a standard eigensolver in pyFBS. A modal damping of $\varepsilon=0.003$ is added to every mode, hence the system is considered as lightly damped. Finally, a complex Gaussian random distributed noise is added to each synthesized acceleration FRF frequency by frequency according to \cref{eq:error}. The parameters defining the noise level in \cref{eq:error} are chosen as $a=1\mathrm{e}{-3},b=4\mathrm{e}{-1},c=2\mathrm{e}{-3},d=1\mathrm{e}{-2}$. Owing to the dynamic characteristics of the system, the noise floor level will be significant for specific input-output combinations (particularly around anti-resonances) while remaining relatively low for others. This corresponds to a realistic scenario and highlights the need for the 'reconstructive' features of the PRF part of the PRANK filter.   

\subsection{Filtering with PRANK}
\label{subsec:NE_2}
\begin{figure} 
	\centering
	\begin{subfigure}{0.495\textwidth}
		\centering
		\includegraphics[width=1\linewidth]{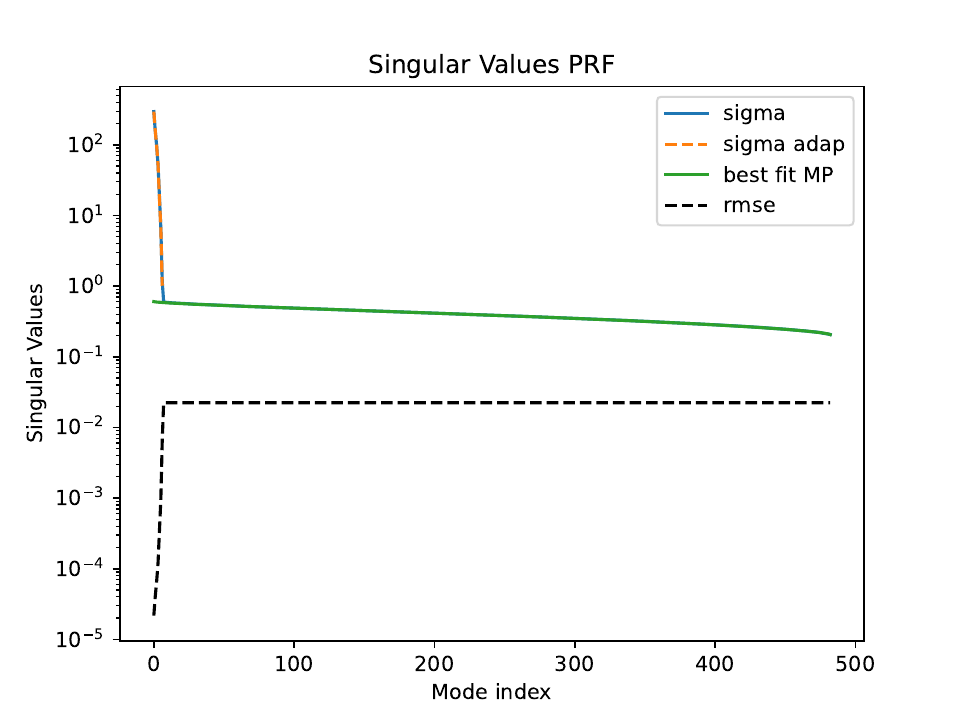}
	\end{subfigure}
	\begin{subfigure}{0.495\textwidth}
		\centering
		\includegraphics[width=1\linewidth]{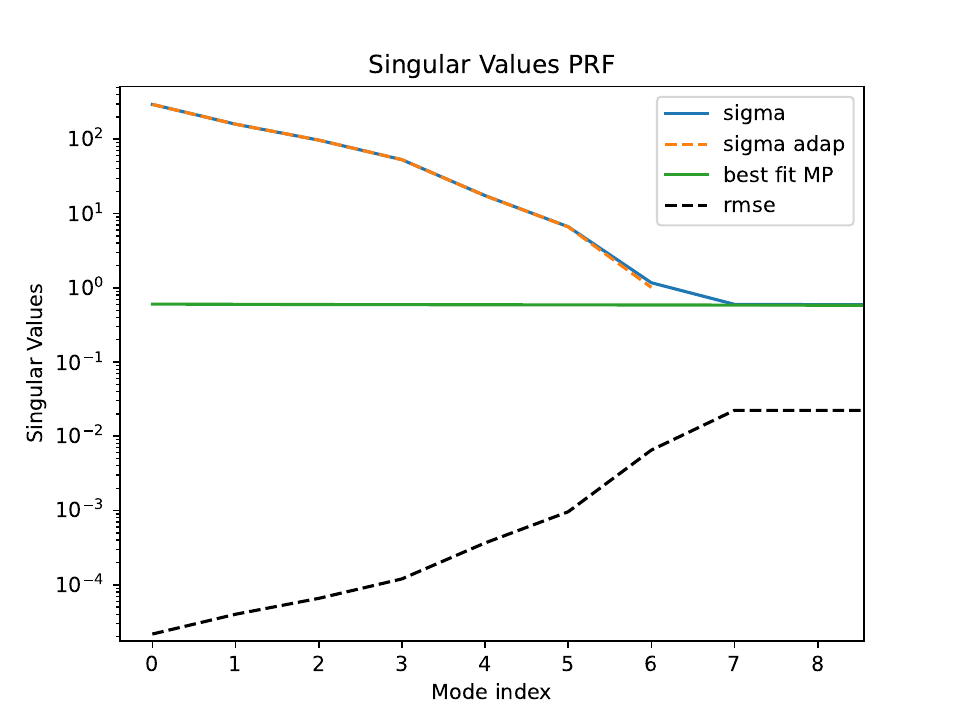}
	\end{subfigure}
	\caption{Use of e15 algorithm in the SVD selection and reconstruction for the PRF stage of a $PRANK_{PH}$ or $PRANK_{HiP}$. Singular value PRF data (blue), best fit Marchenko-Pastur (green), root mean square error per PRF singular value (black), retained and reconstructed singular value data (orange). Left: full SVD. Right: zoom on first 7 modes. }
	\label{fig:e15_HiP_PRF}
\end{figure}

\begin{figure} 
	\centering
	\begin{subfigure}{0.495\textwidth}
		\centering
		\includegraphics[width=1\linewidth]{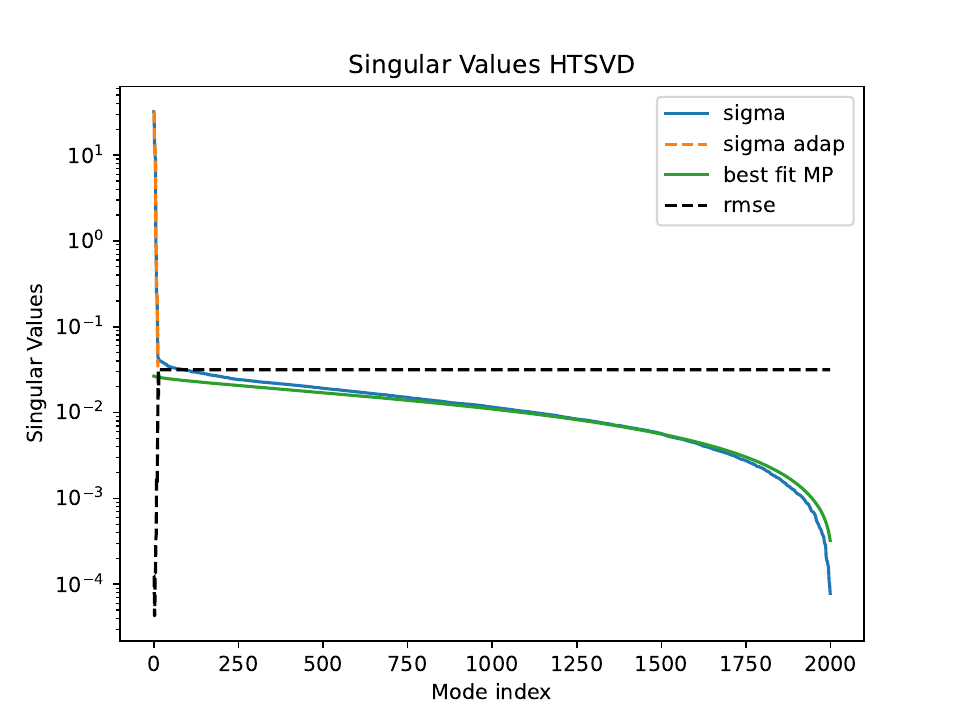}
	\end{subfigure}
	\begin{subfigure}{0.495\textwidth}
		\centering
		\includegraphics[width=1\linewidth]{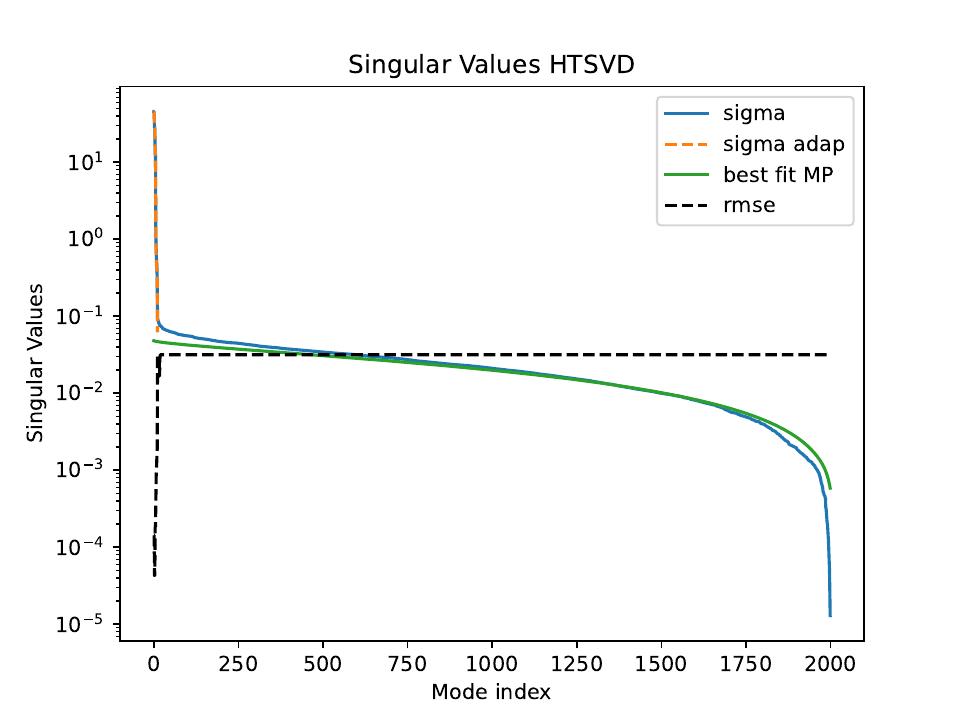}
	\end{subfigure}
	\caption{Use of e15 algorithm in the SVD selection and reconstruction for the Hankel stage of a $PRANK_{PH}$. Singular value Hankel data (blue), best fit Marchenko-Pastur (green), root mean square error per Hankel singular value (black), retained and reconstructed singular value data (orange). $2$ out of $483$ FRFs to be cleaned are shown. }
	\label{fig:e15_PH}
\end{figure}

\begin{figure} 
	\centering
	\begin{subfigure}{0.495\textwidth}
		\centering
		\includegraphics[width=1\linewidth]{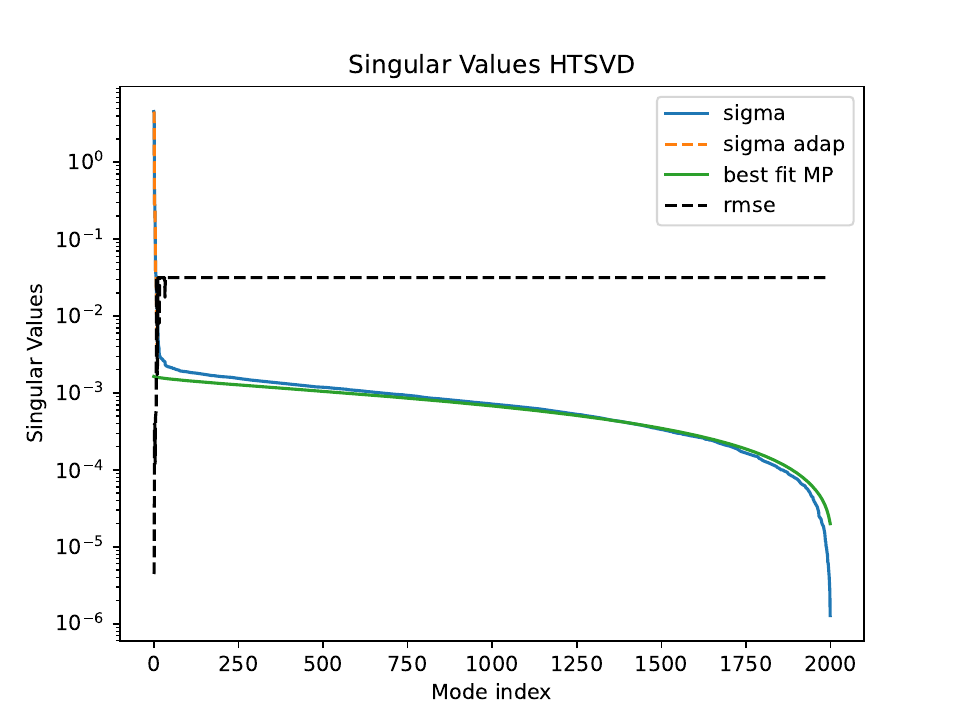}
	\end{subfigure}
	\begin{subfigure}{0.495\textwidth}
	\centering
	\includegraphics[width=1\linewidth]{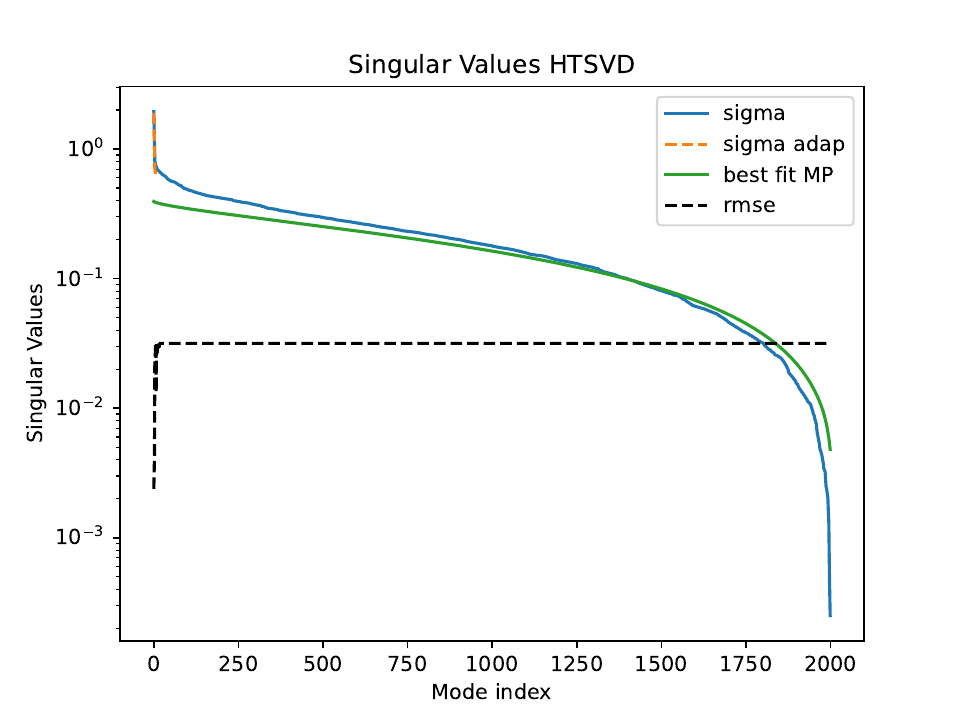}
\end{subfigure}
	\caption{Use of e15 algorithm in the SVD selection and reconstruction for the Hankel stage of a $PRANK_{HiP}$. Singular value Hankel data (blue), best fit Marchenko-Pastur (green), root mean square error per Hankel singular value (black), retained and reconstructed singular value data (orange). $2$ out of $7$ FRFs to be cleaned are shown. Left: Most dominant PRF mode. Right: Least dominant PRF mode.}
	\label{fig:e15_HiP_H}
\end{figure}

\begin{figure} [H]
	\centering
	\begin{subfigure}{0.85\textwidth}
		\centering
		\includegraphics[width=1\linewidth]{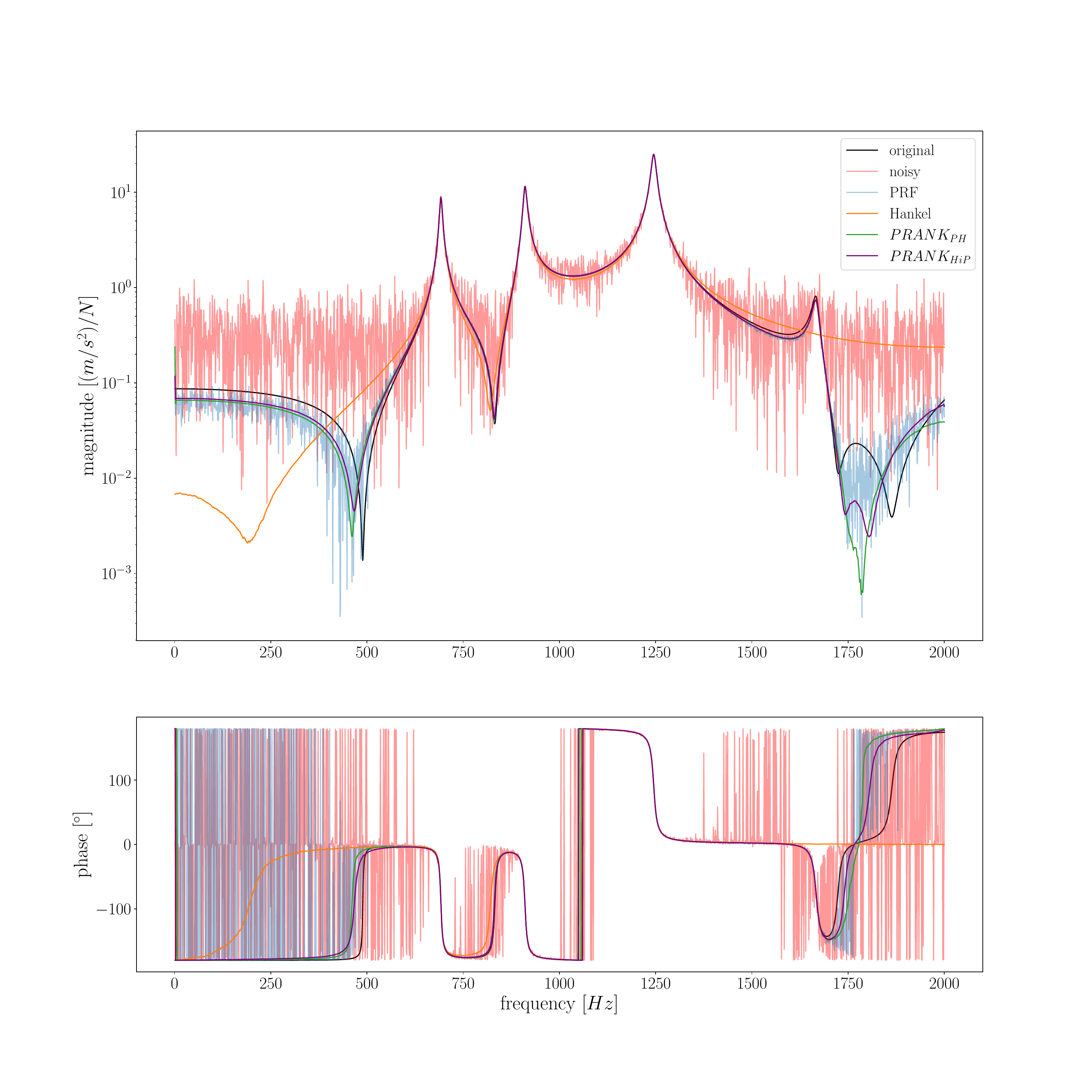}
	\end{subfigure}
	\caption{Comparison between reference (black), noise-polluted (pink), PRF (blue), Hankel (orange), $PRANK_{PH}$ (green), $PRANK_{HiP}$ (purple) data. Magnitude and phase in the frequency range \unit[0-2000]{Hz}. All filters rely on time-based processing and e15 selection with 10\% threshold.}
	\label{fig:num_comp_1}
\end{figure}

\begin{figure} [H]
	\centering
	\begin{subfigure}{1\textwidth} 
		\centering
		\includegraphics[width=0.6\linewidth]{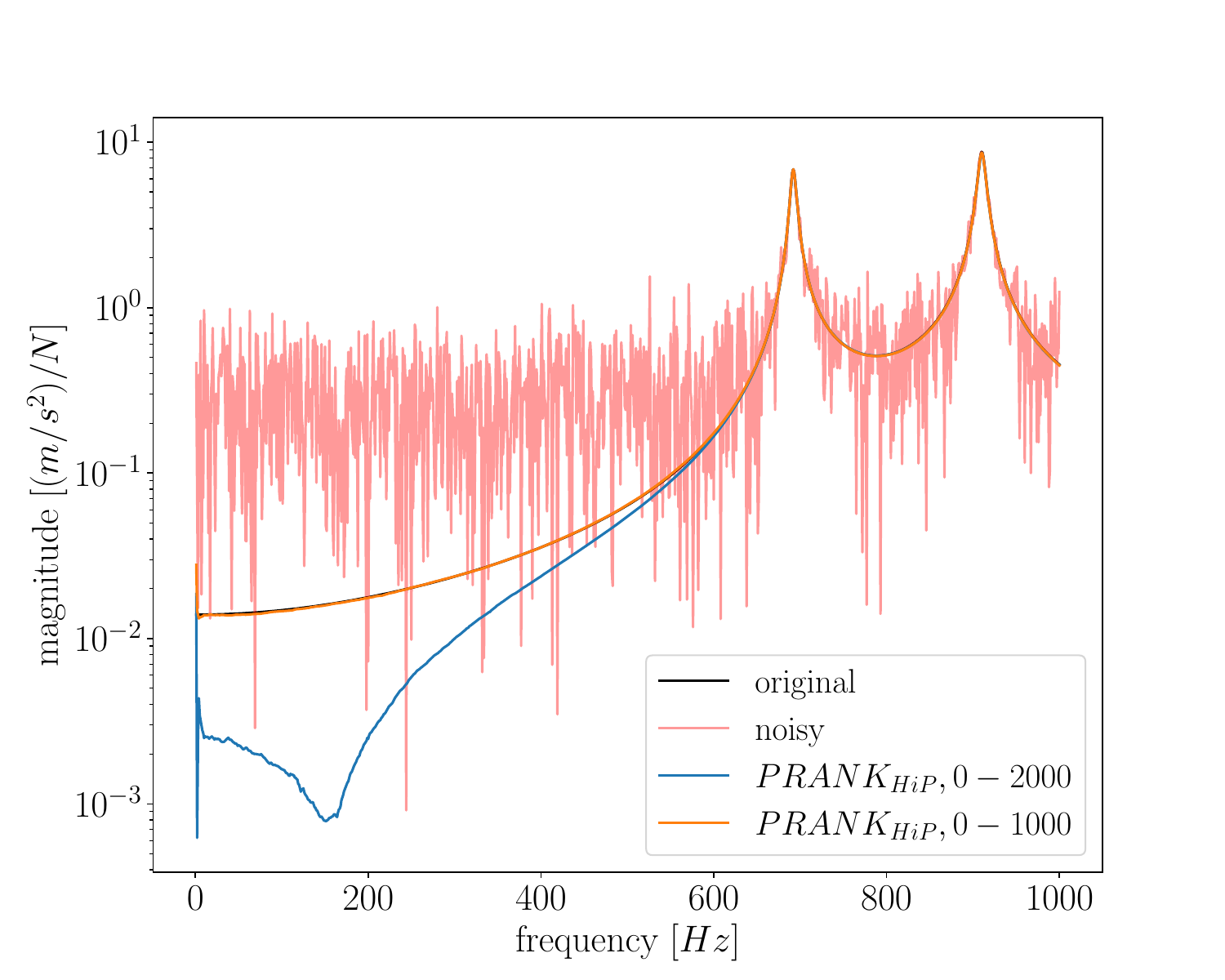}
	\end{subfigure}
	\caption{Comparison between reference (black), noise-polluted (pink), $PRANK_{HiP},0-2000$ (blue), $PRANK_{HiP},0-1000$ (orange) data. The range $0-2000$ or $0-1000$ indicates the frequency range of the FRF dataset on which filtering is applied. Magnitude and phase in the frequency range \unit[0-1000]{Hz}. PRANK relies on time-based processing and e15 selection with 10\% threshold.}
	\label{fig:num_comp_3}
\end{figure}

To compare the performance of different filters, both accuracy and computational costs must be evaluated. Accuracy will be demonstrated by comparing the original (reference), noise-polluted, and PRANK reconstructed data. A consistency indicator is defined as \cite{Haeussler2018}:
\begin{equation}
\label{coh_cr}
\text{consist}([\mathbf{Y}]^{ref}_{n_o \times n_i \times n_k},[\mathbf{Y}]^{filt}_{n_o \times n_i \times n_k})=\dfrac{1}{n_o n_i n_k}\sum_{i=1}^{n_o}\sum_{j=1}^{n_i}\sum_{k=1}^{n_k}\text{coh}(Y_{ij}(\omega_k),Y^{filt}_{ij}(\omega_k))
\end{equation}
where
\begin{equation}
\text{coh}(a,b)=\dfrac{(a+b)(a^H+b^H)}{2(aa^H+bb^H)}
\end{equation}
Computational cost is estimated as the total run time. The computations are made with a PC, with a 3.50 GHz Intel Xeon E3-1240 v5 CPU and 32 GB of RAM. \\
In this section, all filtering results employ the automated e15 selection process. The technique is illustrated in \cref{fig:e15_HiP_PRF}, \cref{fig:e15_PH} \cref{fig:e15_HiP_H} for a time-based PRANK with a tolerance $\mu=10\%$. 
In \cref{fig:e15_HiP_PRF} the PRF SVD selection on the whole dataset is shown. The empirical distribution (green) fits accurately the tail of the singular value curve (blue), thereby representing a good statistical mean to identify a noise floor. Given the computation of a root mean square error (see \cref{sec:PRANK_5}) per mode (black), it appears evident how a few dominant singular values (here the first $7$ values) are sufficiently away from the noise ceiling. The retained and reconstructed singular values are highlighted in orange. \\ In \cref{fig:e15_PH} the Hankel SVD selection within a $PRANK_{PH}$ is demonstrated for a small subset of FRFs ($2$ out of the $483$ to be processed). The shape of the singular value curve, as well the subset of selected dominant values appear similar. This is not true for the Hankel SVD selection within a $PRANK_{HiP}$, illustrated in \cref{fig:e15_HiP_H}. Displayed here are $2$ out of the $7$ dominant left PRF singular vectors. As we move from the primary contribution to the least influential, the level of information retained through the Hankel filtering diminishes. In simpler terms, as we investigate modes in descending relevance, noise significantly corrupts the 'mode' under investigation, necessitating a more robust filtering process.\\
The accuracy and cost of PRANK is summarized in \cref{tab:filter comparison}. 
A time-based formulation is used. PRANK is efficiently removing noise and accurately reconstructing FRFs for all cases. Two considerations follow:
\begin{itemize}
\item The proposed filter $PRANK_{HiP}$ is $45$ times more efficient than the standard (raw) $PRANK_{PH}$ with a negligible difference in accuracy.
\item No relevant differences in accuracy and costs between thresholds $\mu=5\%$ and $\mu=10\%$.
\end{itemize}
A graphic representation of filtering results is given in \cref{fig:num_comp_1}. To highlight the need for PRANK, the filtering action of a PRF TSVD and a Hankel TSVD is also plotted. PRF can detect and reconstruct zeros, albeit with a trade-off in cleaning quality. While Hankel effectively removes noise, it lacks the ability to identify zero locations. On the other hand, both $PRANK_{HiP}$ and $PRANK_{PH}$ yield satisfactory results.\\\\
In \cref{fig:num_comp_3} an important feature of PRANK is demonstrated: The results of filtering are influenced by the frequency range taken into consideration within the processed dataset. Filtering with a limited range of \unit[0-1000]{Hz} proves to be both more efficient (\unit[18]{s} compared to \unit[70]{s}) and more accurate (coherence \unit[0.977]{} versus \unit[0.971]{}) when compared to filtering across an extended range of \unit[0-2000]{Hz}, if the narrower \unit[0-1000]{Hz} range is of interest.\\
As an additional remark, the authors tested the robustness of the filtering process when a reduced/clustered design of experiments is processed (less data available). The idea that an effective PRF operation for dynamics reconstruction requires an extremely diverse input/output spatial distribution is disproved. This finding could be promising for processing techniques, such as dynamic substructuring, that often depend on clustered 'interface' measurements. 

\setlength{\extrarowheight}{2pt}
\setlength{\tabcolsep}{6pt}
\newcolumntype{P}[1]{>{\centering\arraybackslash}p{#1}}

\begin{table}
	\centering
	\small
	\caption{\small{Accuracy and cost of different PRANK options. The coherence is defined with respect to the reference solution according to \cref{coh_cr}. } }
	\begin{tabular}{| m{4cm} | m{3cm} | m{3cm} |}
		\hline
		\textbf{Approach} & \textbf{Coherence [-]} & \textbf{Time [s]} \\
		\hline		
		Noisy  &     0.856      &  -   \\ \hline
		$PRANK_{PH}$, $\mu=10\%$ &  0.985   &	3169	\\		\hline	
		$PRANK_{PH}$, $\mu=5\%$ &  0.986 	& 3188		\\		\hline	
		$PRANK_{HiP}$, $\mu=10\%$ &	0.982 	&  70 \\	\hline	
		$PRANK_{HiP}$, $\mu=5\%$ &	0.982 	&  71 \\	
			\hline		
	\end{tabular}
	\label{tab:filter comparison}
\end{table}

\section{PRANK for an experimental modal analysis with high-speed camera measurements}
\label{sec:EE}

\begin{figure} [h]
	\centering
	\begin{subfigure}{1\textwidth}
		\centering
		\includegraphics[width=1\linewidth]{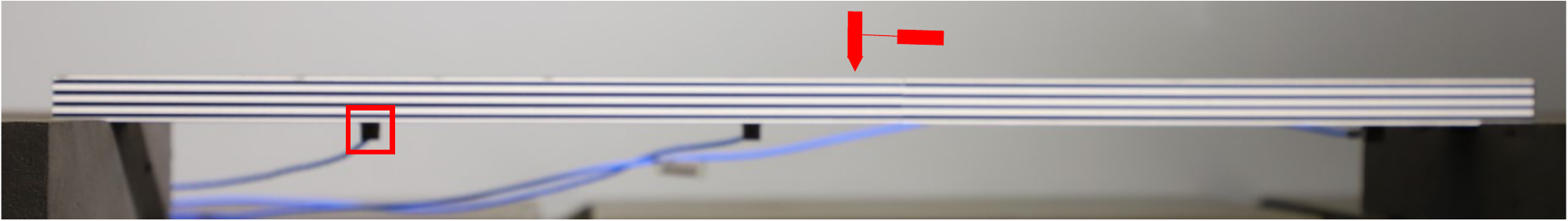}
	\end{subfigure}
	\caption{Beam supported on foam with black-white line pattern for image processing. Focus on impact and reference sensor location (red).}
	\label{fig:beam}
\end{figure}

The filter is tested on high-speed camera displacement measurements, which exhibit significant levels of noise. Camera-based measurements have high potential in modal analysis thanks to an efficient full-field characterization of the structural deformation field. However, the noise-wall often hinders a comprehensive identification of the system dynamics, especially in a high frequency range where displacement amplitudes are typically small. The goal of the campaign is to identify full-field mode shapes at mid- to high-frequency regimes and reconstruct clean FRFs.  \\
The rig consists of a \unit[15x30x500]{mm} steel beam supported at the two ends by foam bricks to represent (quasi) free-floating conditions. The setup is shown in \cref{fig:beam}. A soft impact is performed with a modal hammer on the top of the beam normal to the plane to vibrate the structure freely. A single-head camera, i.e. the Photron Fastcam Nova S6 with \unit[8]{GB} of memory and \unit[1024x1024]{pixel} resolution, is used to capture the 2D full-field motion of the long side of the beam, where a black-white line pattern is attached to provide high contrast and assist the estimation of displacement along the vertical axis. Three \unit[10]{mV/g} triaxial piezo-accelerometers are attached to the bottom of the beam to provide reference measurements (see \cref{fig:beam}) A sampling rate of \unit[20]{kHz} and a measurement time of ca. \unit[1]{s} are employed for both the camera and sensor measurements. 

\subsection{Processing of measurements}
\label{subsec:EE_2}
\begin{figure} [ht]
	\centering
	\begin{subfigure}{1\textwidth}
		\centering
		\includegraphics[width=1\linewidth]{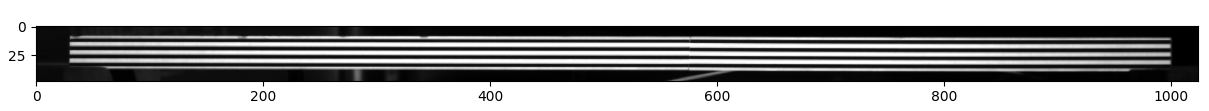}
	\end{subfigure}
	\begin{subfigure}{1\textwidth}
		\centering
		\includegraphics[width=1\linewidth]{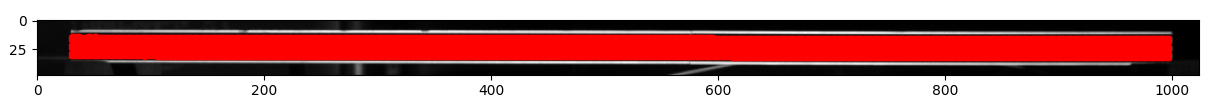}
	\end{subfigure}
	\begin{subfigure}{0.8\textwidth}
		\centering
		\includegraphics[width=1\linewidth]{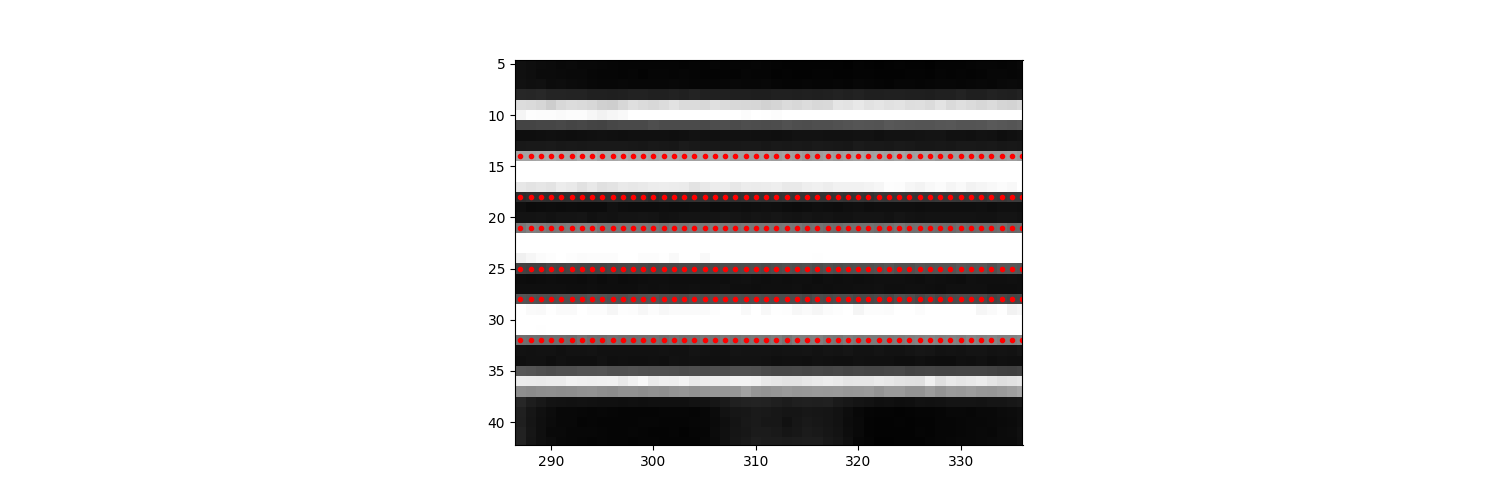}
	\end{subfigure}
	\caption{Image processed with simplified gradient-based optical flow. Top: raw reference image. Center: points selected for displacement identification (red). Bottom: zoom on a small area to highlight the pattern of points selected for displacement identification (red). For each pixel along the horizontal axis, 6 edge-points on the vertical axis are used for an averaged estimation.}
	\label{fig:image}
\end{figure}

\begin{figure} 
	\centering
	\begin{subfigure}{0.495\textwidth}
		\centering
		\includegraphics[width=1\linewidth]{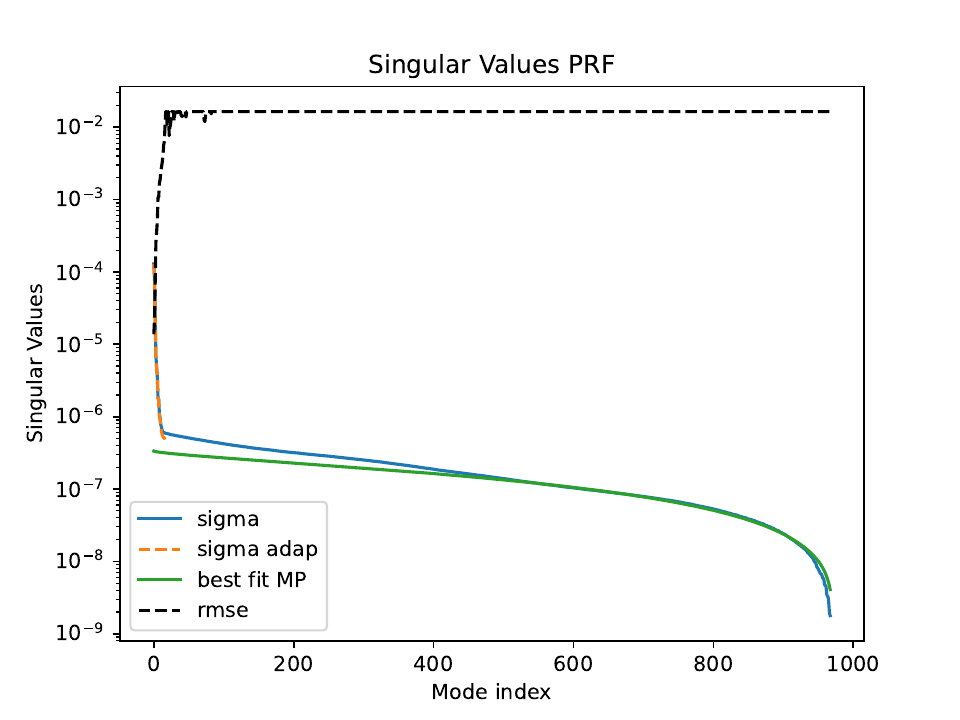}
	\end{subfigure}
	\caption{Use of e15 algorithm in the PRF stage of $PRANK_{HiP}$ for the camera data. Singular value PRF data (blue), best fit Marchenko-Pastur (green), root mean square error per PRF singular value (black), retained and reconstructed singular value data (orange).}
	\label{fig:exp_PRF}
\end{figure}

\begin{figure} 
	\centering
	\begin{subfigure}{0.495\textwidth}
		\centering
		\includegraphics[width=1\linewidth]{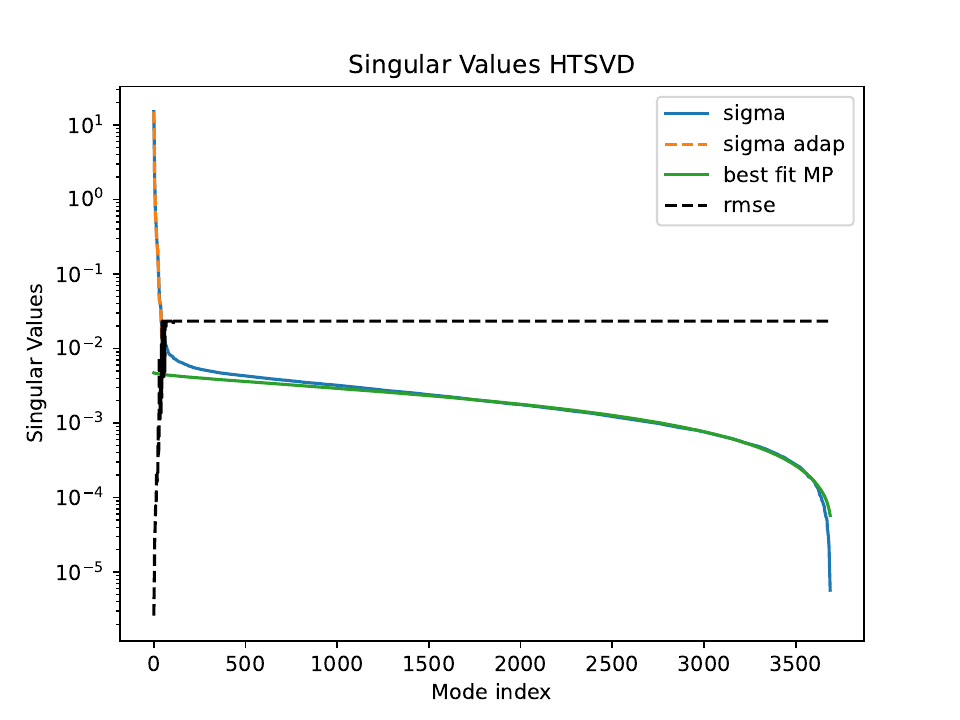}
	\end{subfigure}
	\begin{subfigure}{0.495\textwidth}
		\centering
		\includegraphics[width=1\linewidth]{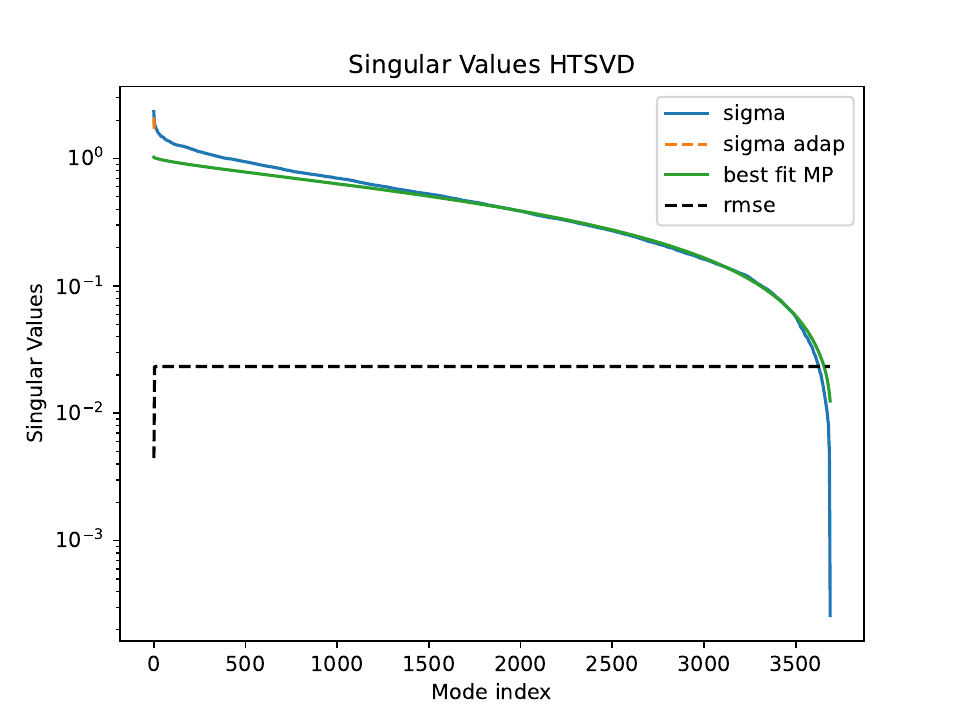}
	\end{subfigure}
	\caption{Use of e15 algorithm in the Hankel stage of $PRANK_{HiP}$ for the camera data. Singular value PRF data (blue), best fit Marchenko-Pastur (green), root mean square error per PRF singular value (black), retained and reconstructed singular value data (orange). Left: Most dominant PRF mode. Right: Least dominant PRF mode.}
	\label{fig:exp_Hankel}
\end{figure}

\begin{figure} 
	\centering
	\begin{subfigure}{0.495\textwidth}
		\centering
		\includegraphics[width=1\linewidth]{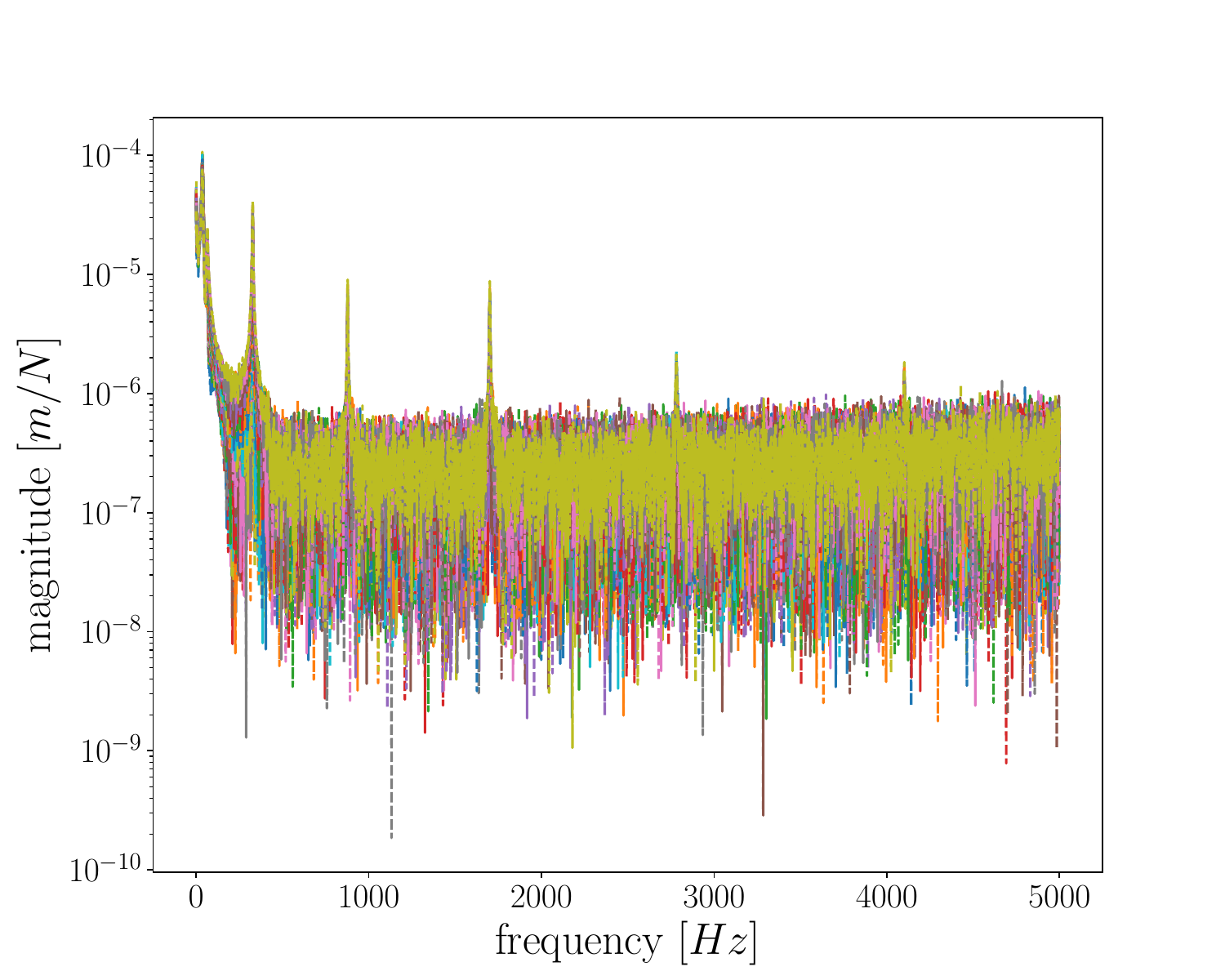}
	\end{subfigure}
	\begin{subfigure}{0.495\textwidth}
		\centering
		\includegraphics[width=1\linewidth]{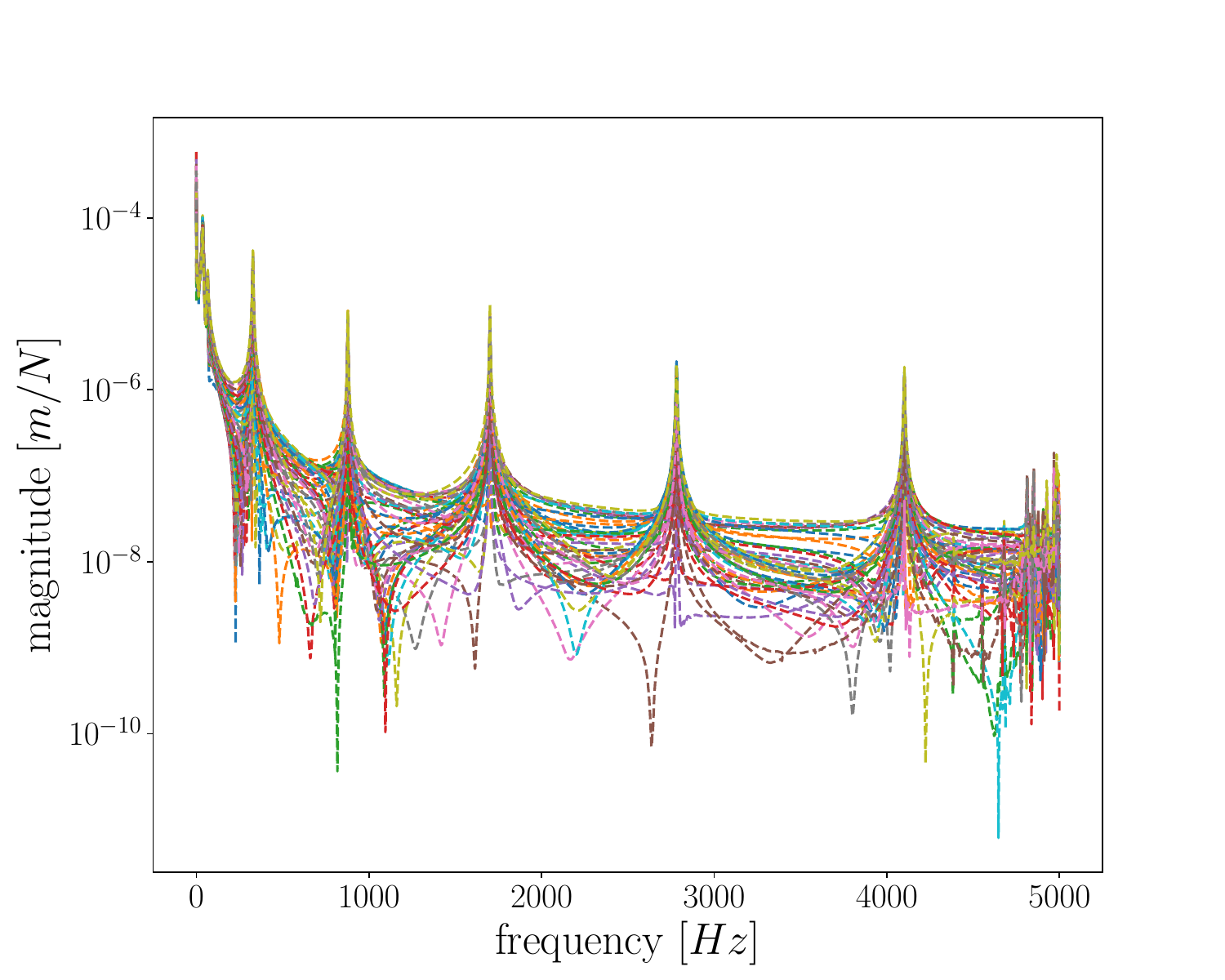}
	\end{subfigure}
	\caption{Overview of processed FRF displacement dataset from high-speed camera data. Only a representative subset ($1/20$) of the entire response dataset is shown for clarity. Left: Unfiltered data. Right: filtered $PRANK_{HiP}$ data.}
	\label{fig:ALL_FRFs}
\end{figure}

\begin{figure} 
	\centering
	\begin{subfigure}{0.8\textwidth}
		\centering
		\includegraphics[width=1\linewidth]{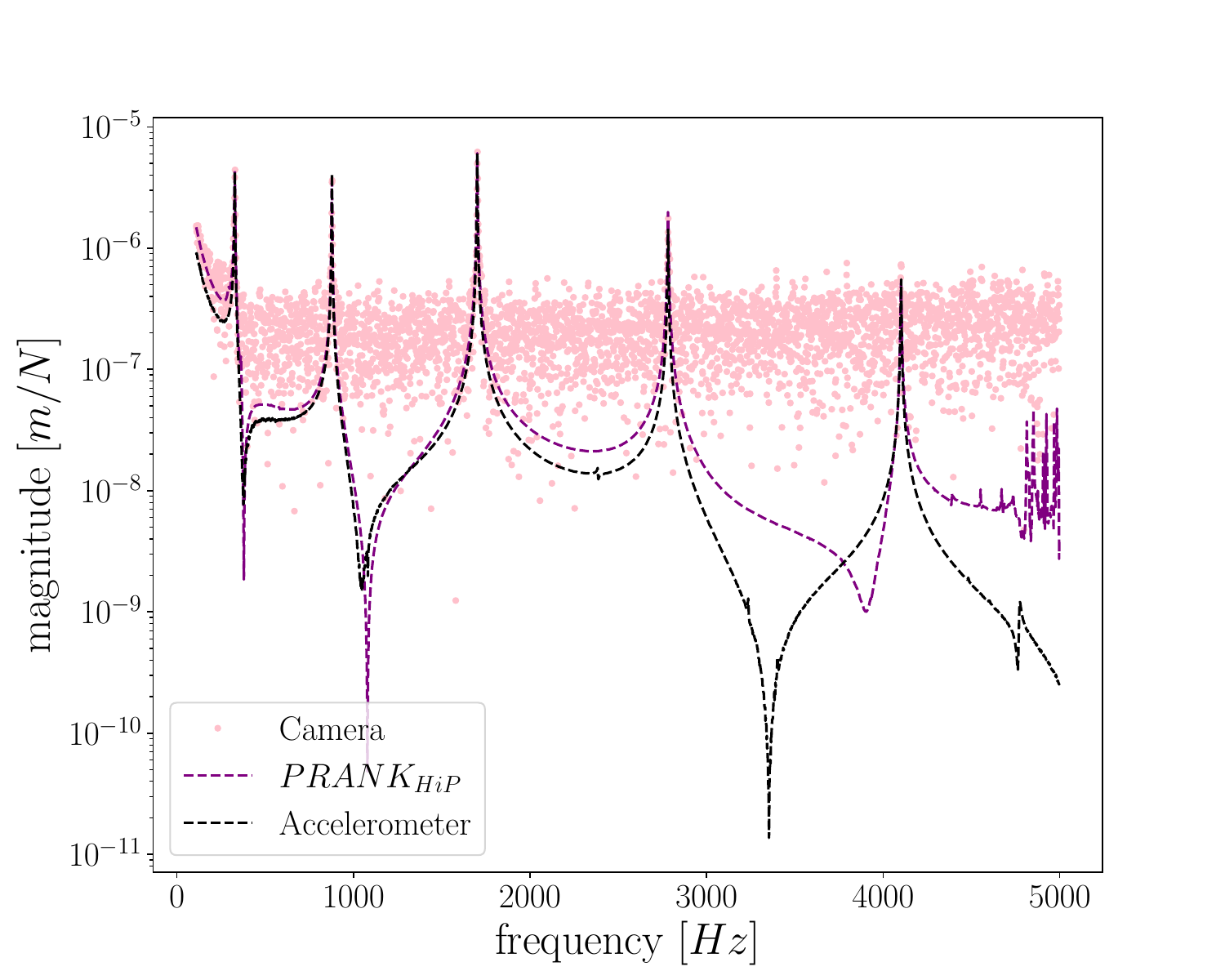}
	\end{subfigure}
	\caption{Comparison between raw camera data (pink), $PRANK_{HiP}$ data (purple) and reference accelerometer data (black) for the response location indicated in \cref{fig:beam}.}
	\label{fig:filtering}
\end{figure}

To identify the motion of the beam from the recorded sequence of images, a Gradient-based Optical Flow approach is chosen. In particular, the Simplified Gradient-based Optical Flow algorithm \cite{Javh2017}, implemented in the free and open-source Python toolbox pyIDI, is employed in this work.\\ 
The points to extract the vertical motion are chosen on each pixel along a horizontal edge of the line pattern from end to end of the entire beam. The identification process is repeated for 6 horizontal edges to get an averaged estimation and decrease the overall amount of noise. The selection of measurement points is highlighted in \cref{fig:image}. \\
A total of $960$ responses are extracted. The estimated displacement is converted from $pixel$ to $mm$ unit by applying a constant scaling factor of ca. $0.5$ \footnote{The distortion due to the distance of the different parts of the beam from the camera lens is considered negligible. Thus, a constant conversion factor is applied to all observed pixels.}, calculated by measuring the size of the lines on the grid. The responses are then synchronized to the accelerometer time signals and cut in time to have a common frequency resolution of ca. \unit[1.35]{Hz}. Frequency response functions are ultimately estimated by using the spectra of the output and input functions.\\
The filter $PRANK_{HiP}$ is applied on the frequency response camera dataset ($968\times1\times3689$, output $\times$ input $\times$ frequency lines, with max. frequency at \unit{5000}{Hz}) with default settings, i.e. time-based filtering with e15 with 10\% threshold for both PRF and Hankel stages. A total of $19$ PRF modes are retained by the algorithm (see \cref{fig:exp_PRF}), which are subsequently cleaned by Hankel (see \cref{fig:exp_Hankel}). The e15 proves to be reliable with real data shaped in the PRF and Hankel data-types, thus confirming its potential for NVH applications. The computations are made with a single-core, 3.50 GHz Intel Xeon E3-1240 v5 CPU and 32 GB RAM. The total run time for the filtering operation is ca. \unit{9}{min}, 9\% of it being associated with the PRF stage and the remaining with the Hankel stage. The amount of data to be processed justifies the use of $PRANK_{HiP}$ because of its efficiency compared with $PRANK_{PH}$/$PRANK_{HP}$. \\ The result of filtering is depicted in \cref{fig:ALL_FRFs} where a subset of unprocessed camera frequency responses are compared with the corresponding filtered ones. The processed data reveal a smooth and clean signal around the resonances and good reconstruction capabilities of the zeros (anti-resonances) that contain the spatial information of the measured dataset. As a side effect of the large filtering action, a small number of responses are not able to restore the peaks covered by the noise wall, and more noticeably, spurious peaks appear at very high frequencies (towards the end of the processed frequency range). \\ To emphasize the high quality of the PRANK reconstruction, the camera data are compared to the reference accelerometer acquisition in \cref{fig:filtering} for the DoF location highlighted in \cref{fig:beam}. Despite the high level of noise, the filter successfully reconstructs the FRF with satisfactory accuracy in both resonance amplitude and anti-resonance location up to \unit[5000]{Hz}. A deterioration of the filtered data quality is visible as frequency increases, as a consequence of the scarcity of the information available above the noise floor.

\subsection{Modal analysis results}
\label{subsec:EE_3}

\begin{figure} 
	\centering
	\begin{subfigure}{0.45\textwidth}
		\centering
		\includegraphics[width=1\linewidth]{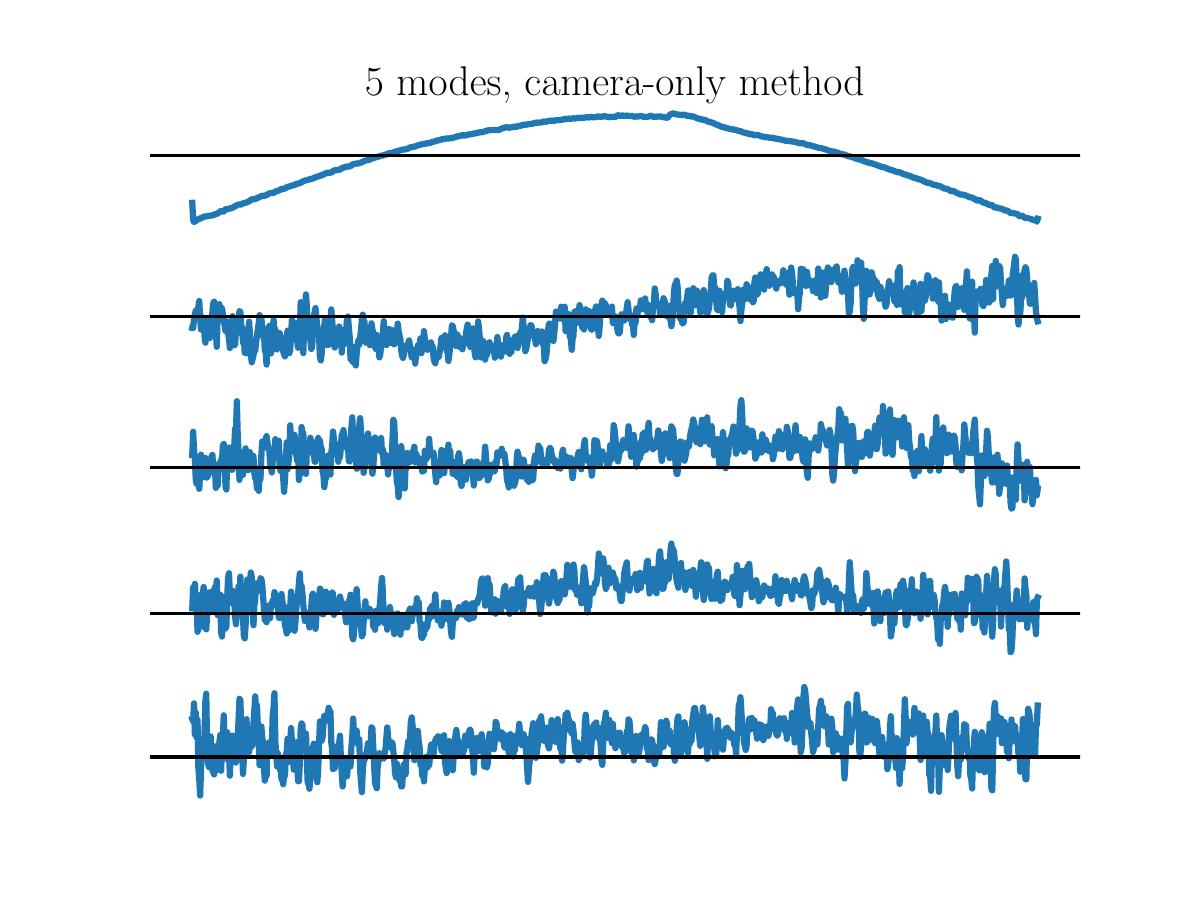}
	\end{subfigure}
	\begin{subfigure}{0.45\textwidth}
		\centering
		\includegraphics[width=1\linewidth]{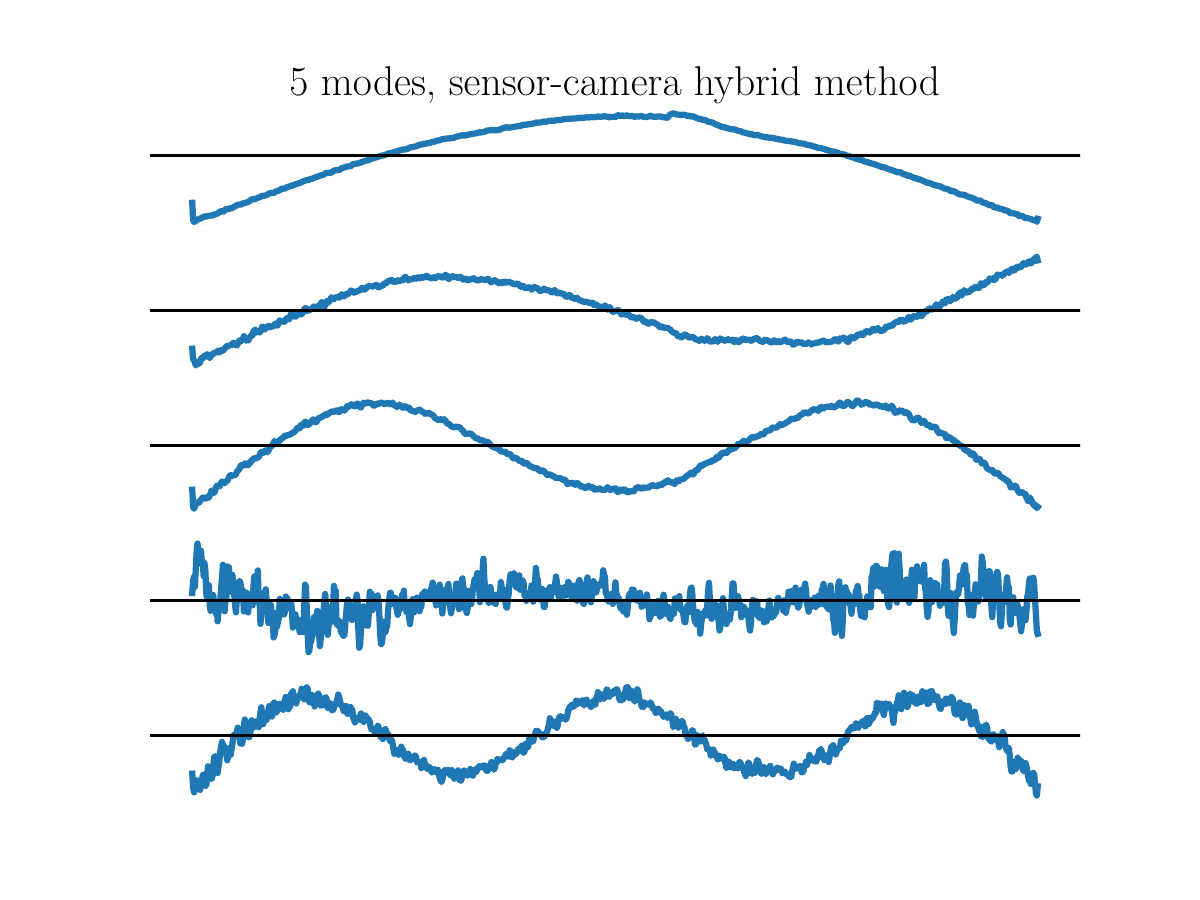}
	\end{subfigure}
	\begin{subfigure}{0.45\textwidth}
		\centering
		\includegraphics[width=1\linewidth]{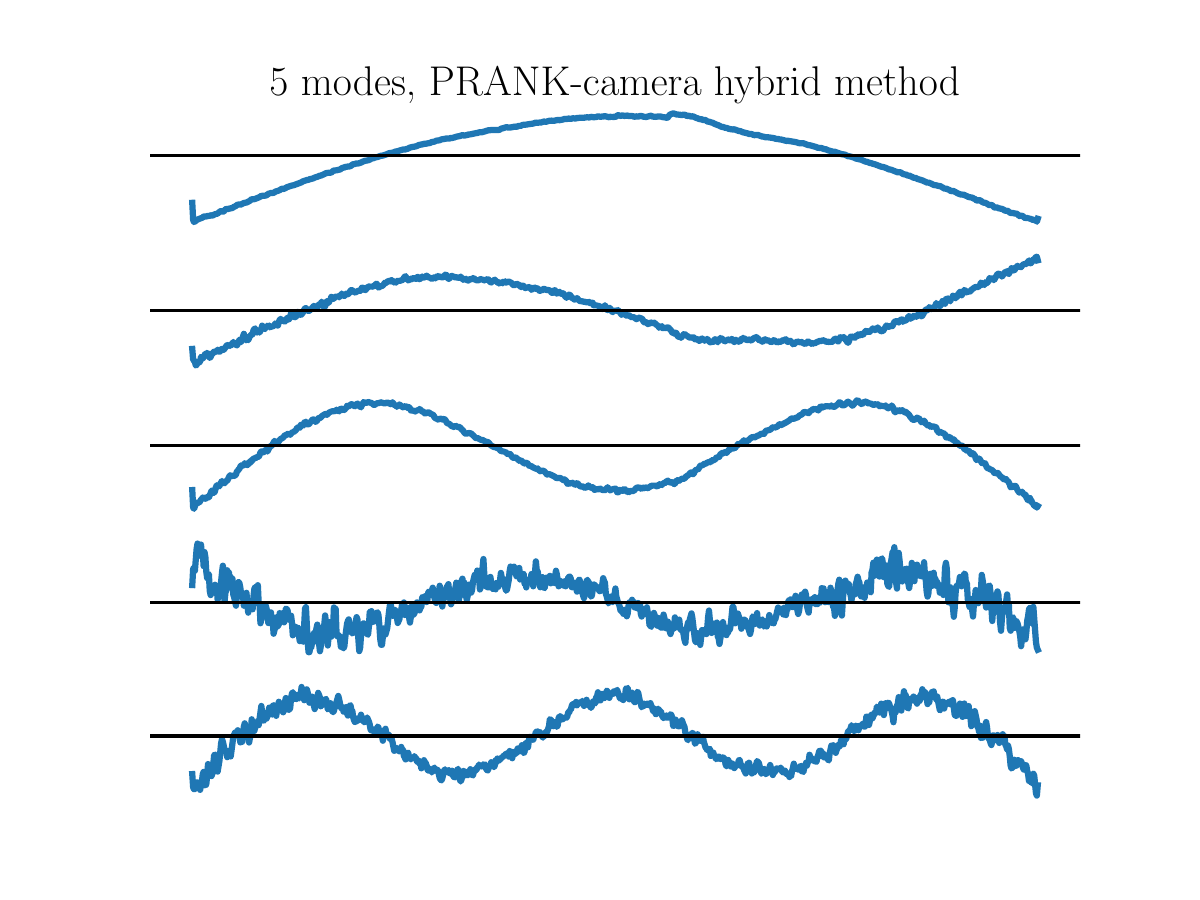}
	\end{subfigure}
	\begin{subfigure}{0.45\textwidth}
		\centering
		\includegraphics[width=1\linewidth]{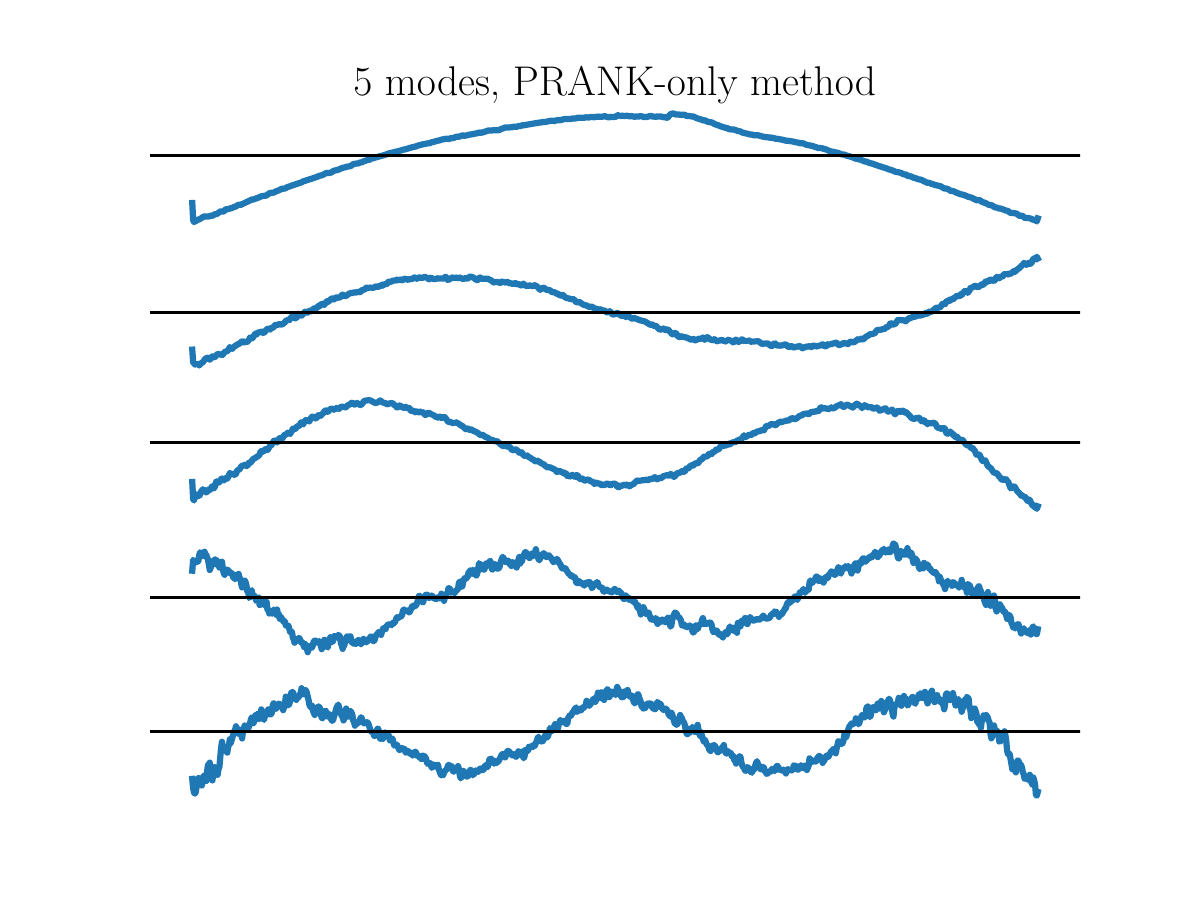}
	\end{subfigure}
	\caption{Comparison between experimental mode shapes estimated with different sets of data and modal identification approaches. LSCF is used for poles determination while LSFD is used for mode shapes and residuals computation. Top-Left: LSCF with raw camera, LSFD with raw camera. Top-Right: LSCF with accelerometer, LSFD with raw camera. Bottom-Left: LSCF with $PRANK_{HiP}$ camera, LSFD with raw camera. Bottom-Right: LSCF with $PRANK_{HiP}$ camera, LSFD with $PRANK_{HiP}$ camera.}
	\label{fig:modes}
\end{figure}

\begin{figure} 
	\centering
	\begin{subfigure}{1\textwidth}
		\centering
		\includegraphics[width=1\linewidth]{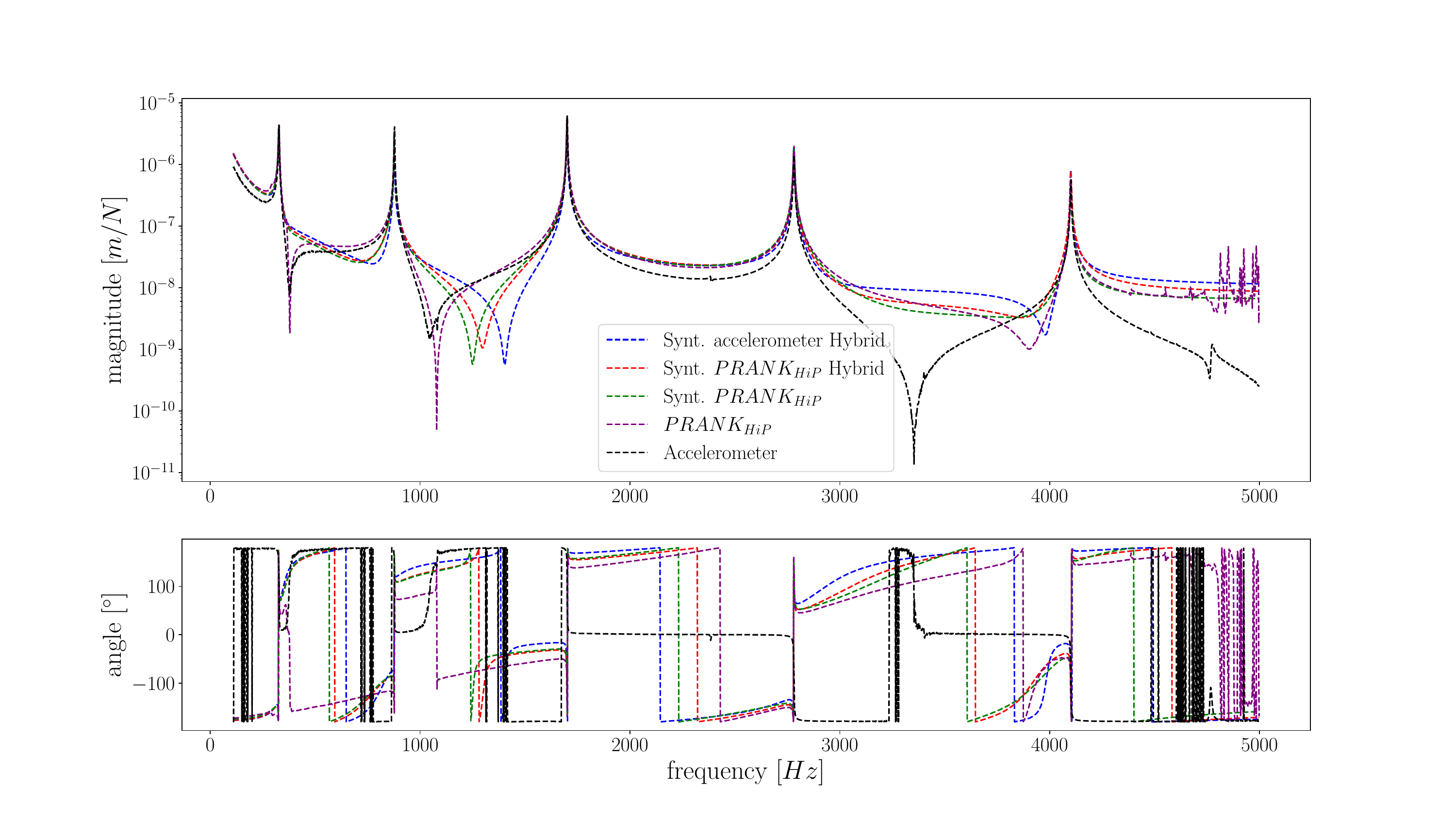}
	\end{subfigure}
	\caption{Comparison between FRFs arising from different sets of data: synthesized accelerometer-camera hybrid data (blue), synthesized $PRANK_{HiP}$-camera hybrid data (red), synthesized $PRANK_{HiP}$ data (green), $PRANK_{HiP}$ data (purple), reference accelerometer data (black). }
	\label{fig:Comparison_FRFs}
\end{figure}

The available data are used to perform an experimental modal analysis. Modal frequencies and damping ratios are extracted via a Least-Square Complex Frequency (LSCF) algorithm, while mode shapes are successively estimated using a Least-Square Frequency Domain formulation (LSFD). 
A hybrid approach is employed to benefit from different types of datasets for the different steps of the modal identification \cite{Javh2018,Bregar2020}. This enables the use of robust localized data to extract the poles/dynamics (e.g. accelerometers) and full-field acquisitions to compute global modal shapes (e.g. camera). All operations are performed with the free and open-source toolbox pyEMA. \\
The poles are manually chosen via a stabilization diagram. 
A total of $4$ different combinations to compute the mode shapes are then compared in \cref{fig:modes}:
\begin{itemize}
\item poles and modes with raw camera data.
\item poles with accelerometer data, modes with raw camera data.
\item poles with filtered camera data, modes with raw camera data.
\item poles and modes with filtered camera data.
\end{itemize}
The unprocessed/raw camera data fail in a proper mode reconstruction, especially at higher frequencies. This can be explained by poor identification of poles due to noise. The filtered camera data show a high-quality modal reconstruction for the tested frequency range with and without employing the unprocessed/raw camera data to calculate modal constants. A potential issue arises near the zeros of the modes when using filtered data in the second stage of modal identification. The reason is that the FRFs at those locations could lack the corresponding dynamical information due to 'overcleaning' during filtering. \\ Overall, the results with the data treated with PRANK appear slightly better than the ones using accelerometers for identifying poles. More importantly, no additional intrusive in-situ equipment (here accelerometers) is needed to ensure a successful modal identification.\\\\
To check the reliability of the identified modal parameters, mode superposition is used to re-synthesize FRFs for all successful application cases (all but the one using raw camera data for poles estimation). Then, the synthesized FRF corresponding to the DoF location highlighted in \cref{fig:beam} is taken for a comparison with the directly filtered camera data and the reference accelerometer response. This is shown in \cref{fig:Comparison_FRFs}. All synthesised FRFs provide similar result. The accuracy of the modal reconstruction is algorithm-specific and will not be further discussed here. The direct non-synthesised filtered FRF is delivering the most accurate response for this case. However, filtering with $PRANK_{HiP}$ does not ensure any physical condition on the response. This aspect will be further discussed in \cref{sec:Discussion}.

\section{Discussion}
\label{sec:Discussion}
PRANK is the idea of combining multiple flavors of TSVD based on the topological representation of the available measured MIMO spectral dataset. The filtering algorithm is incorporated within a streamlined and automated framework. The results arising from the analytical example in \cref{sec:SVR}, the numerical case in \cref{sec:NE} and the experimental application in \cref{sec:EE} display the potential of PRANK in the field of vibration analysis. PRANK focuses on the removal of random error and outliers. No systematic error is treated. Since the core of the filtering is based on a least square TSVD estimator, in a realistic scenario, namely when the conditions assumed in \cref{sec:PRANK_sec} are not fully satisfied, some distortion of the desired underlying signal will inevitably occur. For that reason, it is strongly recommended to apply PRANK on the measured data only if necessary, i.e. a 'relevant' amount of noise is visible. The interpretation of 'relevant' is up to the user and will depend on the specific application.\\
At the light of the experimental application of \cref{sec:EE}, it is worth to clarify the difference between PRANK, i.e. a SVD strategy, and a classic modal decomposition approach. Both decompose the dataset in a simple linear combination of fundamental contributions. If a modal approach seeks to find mass and stiffness orthogonal mode shapes, the singular approach retrieves orthonormal singular vectors. The information contained in the two bases is generally the same, although distributed differently. Due to its mathematical roots, PRANK is not capable of preserving physical properties of a linear time-invariant system such as passivity (see \cref{sec:PRANK_sec}) and reciprocity. Thus, if the goal is to estimate modal parameters, PRANK is not an alternative to modal identification, but rather a supplementary function (see results in \cref{fig:modes}). In this context, it could be an additional option to other existing noise removal techniques associated with modal identification (e.g. the Maximum Likelihood Modal Model estimator for Polymax \cite{El-Kafafy2013}). If, instead, the goal is to clean a dataset while preserving the dominant underlying dynamics, PRANK may be a robust alternative to a modal reconstruction/synthesis (see results in \cref{fig:Comparison_FRFs}). Further study of PRANK in combination with or as an alternative to modal identification is envisioned.\\
The potential enhancement of PRANK with physicality-preserving constraints is under research. The application of PRANK within NVH operations such as Frequency-Based Substructuring (FBS) and Transfer Path Analysis (TPA) is left for future work.

\section{Conclusion}
\label{sec:Conclusion}
Noise and disturbances arising from experimental measurements are often unavoidable and, if not adequately treated, can lead to erroneous or incomplete findings.\\
This article presents PRANK: A robust, efficient and automated algorithm based on singular values truncation designed to remove noise and outliers from measured MIMO response datasets. The approach capitalizes on the complementary effects of PRF (to detect outliers) and of Hankel (to remove random noise) SVD truncation mechanisms. The sequential use of PRF followed by Hankel, termed as $PRANK_{PH}$, demonstrates superior effectiveness compared to employing each filtering technique independently. A novel way of mixing the two techniques, $PRANK_{HiP}$, is formulated that goes beyond a sequential application, resulting in comparable cleaning effectiveness, significantly reduced computational cost, and enhanced versatility. The selection of the truncation basis is automated through the integration of the e15 algorithm proposed in literature, which is proven applicable to NVH datasets in the form used by PRANK. An analytical example is used to show the applicability, benefits and drawbacks of different SVD truncation approaches. A numerical example shows the efficacy of the full PRANK functionality, highlighting accuracy and cost of the proposed strategy. Excellent noise removal and signal reconstruction is demonstrated in the context of a full-field camera-based experimental modal analysis. \\
Future research will be made to extend the usability and reliability of the algorithm and to investigate applications to different fields and problems.

\section*{Conflict of Interest}

On behalf of all authors, the corresponding author states that there is no conflict of interest.

\appendix%

\printbibliography

\end{document}